\documentclass[journal]{new-aiaa}
\usepackage[utf8]{inputenc}
\usepackage{textcomp}

\usepackage{graphicx}
\usepackage{subcaption}
\usepackage{amsmath}
\usepackage{bm}
\usepackage{siunitx}
\usepackage{longtable,tabularx}
\title{Optimal TRACON Descent Procedures under Wind Uncertainty and Fuel Savings Factors}

\author{Yutian Pang\footnote{Postdoctoral Fellow, Department of Aerospace Engineering and Engineering Mechanics, AIAA Member.} and John-Paul Clarke\footnote{Professor, Department of Aerospace Engineering and Engineering Mechanics. AIAA Fellow.}}
\affil{The University of Texas at Austin, Austin, Texas, 78712}

\begin{document}

\maketitle

\begin{abstract}

A terminal-area descent procedure need to perform across the wind climatology rather than a single wind condition. Although flight demonstrations of the delayed deceleration approach (DDA) showed substantial fuel savings, DDA combined late deceleration with a steeper $3.77^\circ$ final descent, obscuring the contribution of each design choice. In this work, we propose the continuous-descent delayed deceleration approach (CDDA), which applies delayed deceleration to a continuous descent approach (CDA) profile without a level segment before glideslope intercept. A simulation-based stochastic optimization selects flap deployment trigger speeds and glideslope-capture distance to minimize expected fuel under wind uncertainty subject to a given stabilized-approach probability. An optimal control reduction limits the design space to a few hundred candidates, enabling exact expectation over a weighted wind grid using six-degree-of-freedom fast-time simulations. CDA and CDDA are optimized at matched final angles of $3.00^\circ$, $3.50^\circ$, and $3.77^\circ$ for the A319, B737-800, B767-400, and A340-300. Results show that deceleration architecture is the weakest factor, yielding 0.3--3.9\% savings at $3.50^\circ$ and becoming material only for the B767-400 at $3.77^\circ$. Glideslope angle dominates fuel saving. At the $3.50^\circ$ Category D design maximum, optimized CDDA reduces expected fuel by 11-21\% relative to optimized $3^\circ$ CDA, while flap-schedule optimization adds 2-17\%. The DDA level segment acts primarily as a tailwind-robustness buffer, and the $3.77^\circ$ final exceeds the 1,000 ft/min stabilized-approach sink-rate element, limiting its near-term operational applicability.
\end{abstract}

\section*{Nomenclature}

{\renewcommand\arraystretch{1.0}
\noindent\begin{longtable*}{@{}l @{\quad=\quad} l@{}}
$a_j, b_j$ & endpoints of the $j$th unstabilized wind interval, kt \\
$d$ & along-track distance to the runway threshold, nautical mile \\
$d_{\mathrm{cap}}$ & glideslope-capture distance from the runway threshold, nautical mile \\
$d_i^{\mathrm{bs}}, d_i^{\mathrm{plan}}$ & backstop and plan-crossing deployment distances of flap detent $i$, nautical mile \\
$d_r, h_r^{\mathrm{floor}}$ & distance and altitude floor of published crossing restriction $r$ \\
$D$ & aerodynamic drag \\
$f(\bm{x}, w)$ & fuel burned from the corridor entry fix to the runway threshold, kg \\
$g$ & gravitational acceleration \\
$G$ & number of flap-detent groups \\
$h_a$ & wind-profile anchor altitude (10{,}000 ft), ft \\
$h_g$ & gate height of the near-surface wind freeze (1{,}000 ft), ft \\
$h_{\mathrm{level}}$ & glideslope-capture altitude, ft \\
$h_{\mathrm{rwy}}$ & field elevation at the runway, ft \\
$h^{\mathrm{plan}}(d; \bm{x})$ & altitude of the zero-wind plan at distance $d$, ft \\
$J$ & number of unstabilized wind intervals \\
$K$ & number of wind-quadrature nodes \\
$m$ & aircraft mass \\
$\dot m_f$ & fuel-flow rate \\
$n_x$ & longitudinal load factor \\
$q(\bm{x})$ & realized design: capture distance and trigger ladder \\
$R$ & number of published crossing restrictions \\
$s(\bm{x}, w)$ & stabilized-approach indicator (1 if stabilized, 0 otherwise) \\
$t_{\mathrm{entry}}, t_{\mathrm{gate}}, t_{\mathrm{thr}}$ & corridor-entry, 1{,}000 ft gate, and threshold crossing times, s \\
$T$ & thrust \\
$V_{\mathrm{CAS}}$ & calibrated airspeed, kt \\
$V_i$ & deployment trigger calibrated airspeed of flap detent $i$, kt \\
$V_{\min,g}, V_{\max,g}$ & placard (minimum, maximum) speeds of flap group $g$, kt \\
$V_{\mathrm{REF}}$ & landing reference speed, kt \\
$V_T$ & true airspeed, kt \\
$w$ & along-corridor wind component at $h_a$, kt (tailwind positive) \\
$W$ & wind random variable, kt \\
$W(h;w)$ & wind speed at altitude $h$, kt \\
$\bm{x}$ & procedure design vector \\
$x_g$ & trigger-speed offset of flap group $g$, kt \\
$\mathcal{D}$ & admissible set of capture distances, nautical mile \\
$\mathcal{F}(\bm{x})$ & unstabilized (failure) wind set of design $\bm{x}$ \\
$\mathcal{X}$ & design set \\
$\gamma$ & flight-path angle \\
$\gamma_f$ & final-descent (glideslope) angle, deg \\
$\delta, \delta_{\mathrm{land}}$ & flap-gear configuration and landing configuration \\
$\Delta w$ & wind-quadrature node spacing, kt \\
$\varepsilon$ & allowed probability of an unstabilized approach \\
$\eta_g$ & half-width of the placard window of flap group $g$, kt \\
$\Lambda$ & lattice of realized designs \\
$\mu_g$ & midpoint of the placard window of flap group $g$, kt \\
$\sigma_w$ & wind standard deviation, kt \\
$\tau_i$ & deployment time of flap detent $i$, s \\
$\varphi_{\mathrm{TN}}$ & truncated-normal wind density \\
$\omega_k$ & quadrature weight of wind node $w_k$ \\
\multicolumn{2}{@{}l}{Subscripts}\\
$g$ & flap group index \\
$i$ & flap detent index \\
$j$ & unstabilized-interval index \\
$k$ & wind node index
\end{longtable*}}
\setcounter{table}{0}

\section{Introduction}
\lettrine{A}{t} the busiest airports in the National Airspace System, runway capacity and terminal maneuvering area (TMA) operations constrain throughput. A large share of flight delay, fuel burn, and noise exposure also occurs. When arrival demand exceeds runway capacity, controllers in the terminal radar approach control (TRACON) facility absorb the excess through tactical vectoring, path stretching, and holding. These actions lengthen low-altitude flight, where fuel flow and community noise are especially sensitive to how the aircraft is flown. Terminal arrival automation dates to the Center-TRACON Automation System, where Traffic Management Advisor, Descent Advisor, and Final Approach Spacing Tool were developed and flight-tested at NASA Ames in the 1990s \cite{denery1997ctas}. Subsequent work on sequencing, scheduling, and machine-assisted vectoring, reviewed in Sec.~II, follows the same direction as trajectory-based operations (TBO) \cite{torres2012tbo}, where trajectories are planned in advance rather than managed clearance by clearance.

The prior work developed the horizontal component of this automation framework \cite{pang2026trajectory}. Arrivals are sequenced and scheduled on a rolling horizon, and each aircraft receives an implementable lateral path with a segment-wise commanded speed profile. However, in this work, fuel consumption was evaluated on a simplified piecewise vertical profile. That simplification is adequate for ranking sequencing policies, but it excludes the vertical choices that strongly affect terminal-area fuel burn, where descent begins, where deceleration occurs, and when high-lift devices are extended \cite{dumont2012fuel}. The present paper supplies this vertical component. Unlike a tactical lateral clearance, in real-world air traffic control practice, a vertical procedure is published, coded into the flight management system (FMS), and flown repeatedly. The design must therefore be selected once and evaluated over the wind conditions in which it will operate.

Two aircraft descent architectures motivate the study. The continuous descent approach (CDA) reduces low-altitude level flight by descending at or near idle thrust from the top of descent. Its fuel and noise benefits have been established by design studies and flight tests \cite{clarke2004continuous}, and the concept has entered charted practice as the optimized profile descent \cite{clarke2013optimized}. On the other hand, the delayed deceleration approach (DDA) extends the fuel-saving principle to the speed profile, where the aircraft remains clean at its descent speed as long as operationally possible, reducing drag, thrust, and fuel burn over a larger fraction of the arrival \cite{dumont2012fuel, sandberg2016delayed}. Thomas and Hansman \cite{thomas2021modeling} modeled DDA speed profiles for current aircraft and demonstrated a steep-final DDA variant during the 2019 ecoDemonstrator flight-test program. That demonstration delayed deceleration and steepened the final from $3^\circ$ to $3.77^\circ$ simultaneously, so the measured benefit does not identify the separate contributions of deceleration architecture and glidepath angle. The flap-deployment schedule is a third candidate driver because flight data link fuel burn directly to airspeed and time in high-lift configuration \cite{dumont2012fuel}.

In practice, a published descent procedure cannot be re-optimized for each day's wind. It is a fixed design operated under uncertain conditions. The design problem is therefore stochastic, choose the procedure parameters that minimize expected fuel over the wind climatology while satisfying a high-probability stabilized-approach requirement. Kendall and Clarke \cite{kendall2020stochastic} used the same design philosophy for area navigation noise-abatement procedures. Here the objective is the expected fuel, and the operational requirement is expressed as a chance constraint. Deterministic trajectory optimization establishes what is achievable in a known wind, where published-procedure design requires optimization over the wind distribution.

The study evaluates three fuel-saving factors in terminal-area final descent, (i) deceleration architecture, (ii) glideslope angle, and (iii) flap-deployment schedule. The first two are properties of the published procedure, and the third is optimized within each procedure. Both architectures are given the same design freedom, wind uncertainty, and stabilization requirement, and they are compared at matched final angles. This structure isolates architecture effects from glidepath effects and baseline tuning. The case study is set as the northwest transition to Runway 8L at Atlanta, described in Sec.~III.A. The method designs the published procedure rather than a controller advisory or onboard assistance function. Here, we highlight the major contributions of this research work as,
\begin{itemize}
\item A continuous-descent delayed deceleration approach (CDDA), which places DDA-style late deceleration on the continuous-descent profile of the CDA.
\item A chance-constrained stochastic program that selects flap-deployment trigger speeds and glideslope-capture distance to minimize expected fuel subject to a 95\% stabilized-approach probability \cite{fsf2009stabilized}.
\item A finite-lattice solution method in which an optimal-control reduction leaves a few hundred candidates, enabling weighted-grid evaluation without Monte Carlo noise or surrogate modeling.
\item A factorial comparison of CDA and CDDA at matched $3.00^\circ$, $3.50^\circ$, and $3.77^\circ$ glideslope finals for four different airframes (two narrow and two wide bodies).
\item Quantification of the DDA level segment as a tailwind-robustness buffer and of the 1 kt wind grid needed to verify the chance constraint.
\end{itemize}

The factorial shows that matched-angle CDA and CDDA designs burn similar expected fuel. Delayed deceleration contributes 0.3--3.9\% at the compliant $3.50^\circ$ final and is material only for the B767-400 at $3.77^\circ$. Glideslope angle shows the larger effect, within the CDA alone, the $3.50^\circ$ Category D maximum saves 8--18\% relative to $3.00^\circ$, and the $3.77^\circ$ Category C maximum saves 12--22\%. The optimized flap schedule adds 2--17\% relative to fixed rules. Section~V develops these results, including stabilization-risk use and the airframe-dependent value of the level deceleration segment.

In the rest of the paper, Section~II reviews terminal-area automation, continuous descent operations, delayed deceleration procedures, and procedure design under uncertainty. Section~III describes the arrival corridor, procedure architectures, fast-time simulation, and wind model. Section~IV states the stochastic optimization problem and search procedure. Section~V presents the results, and Sec.~VI concludes.

\section{Related Work}

\subsection{Terminal-Area Automation and Arrival Scheduling}
Automated arrival management has been studied at every control layer between the en route boundary and the runway. At the runway itself, sequencing and scheduling under wake-vortex separation is a classical operations research problem, surveyed by Bennell et al. \cite{bennell2011runway}. Balakrishnan and Chandran \cite{balakrishnan2010scheduling} gave scalable dynamic-programming algorithms for runway scheduling under constrained position shifting, which bounds how far any aircraft may move from its first-come-first-served position, and Anagnostakis and Clarke \cite{anagnostakis2003runway} decomposed mixed arrival and departure runway planning into two stages chosen so that the first depends only on quantities that stay predictable under uncertainty. Later formulations widened the scope from the threshold to the terminal airspace: Desai and Prakash \cite{desai2016terminal} found that sequencing over the full fix-and-route structure of the TMA reduces delay well beyond what runway-only models achieve, Ng et al. \cite{ng2024tasp} coupled runway assignment, speed control, holding, and point-merge decisions in a matheuristic validated on Singapore Changi traffic, and Gui et al. \cite{gui2025metaheuristic} applied rolling-horizon metaheuristics to the same problem class. The vectoring itself has also been automated. Dhief et al. \cite{dhief2023delayabsorption, dhief2025vectoring} generate conflict-free vector geometries that absorb assigned delays before the final approach fix, Liang et al. \cite{liang2017parallel} integrated a multilevel point-merge structure with continuous-descent-style profiles for parallel runways at Beijing Capital, and Diffenderfer et al. \cite{diffenderfer2013integration} demonstrated a TRACON tool that adjusts arrival spacing dynamically to the departure queue on a dependent runway. Empirical analysis of inbound flows also shows that appreciable arrival-time uncertainty persists from the en route phase to the top of descent \cite{lubig2025uncertainty}.

These systems decide which aircraft lands when, and along which lateral path. The vertical realization of each arrival is taken as given, typically as a nominal descent profile per aircraft type, and our horizontal framework \cite{pang2026trajectory} makes the same simplification. What an aircraft does in the vertical plane is governed instead by the published arrival and approach procedure, which must be designed before any tactical automation runs. How such procedures have been designed, and what their design leaves open, occupies the rest of this section.

\subsection{Continuous Descent Operations}
The CDA investigation begins with the effort on reducing community noise. Clarke \cite{clarke1997systems} showed with a combined flight-simulation, noise, and land-use analysis environment that a decelerating, low-thrust approach reduces single-event noise while remaining flyable and acceptable to pilots. The Louisville design study and flight test \cite{clarke2004continuous} established the operational case, with fuel savings of roughly 400 to 500 lb per flight and peak-noise reductions of several decibels. Publicly charted optimized profile descents followed, first at Los Angeles, where arrivals on the new procedure saved an average of 25 gallons of fuel each \cite{clarke2013optimized}. Wider use has been limited by predictability: idle-thrust descents vary with aircraft type, weight, and wind, so controllers often add separation buffers that reduce throughput \cite{park2015optimal}. Ren and Clarke \cite{ren2007modeling, ren2007separation} addressed this problem with the Tool for the Analysis of Separation and Throughput (TASAT), a Monte Carlo fast-time simulation that sets metering-fix separations to a specified confidence level. Atlanta flight trials showed that manually delivering aircraft to those separations is itself difficult \cite{lowther2008enroute}, and a New York metroplex study found CDA viable as standard operation when a scheduler makes the descents conflict free \cite{cao2011evaluation}.

Optimization of the CDA profile itself closely followed. Park and Clarke \cite{park2015optimal} posed the minimum-fuel and minimum-time vertical profiles as a multiphase optimal control problem, computed performance bounds for a light and a heavy airframe, and showed that suboptimal profiles assembled from FMS-flyable vertical navigation modes perform nearly as well as the true optimum. Their extension to descents in wind \cite{park2016vertical} characterized the optimal arcs analytically and generated them by backward and forward integration at negligible computational cost. Multiobjective trajectory generation for integrated avionics and air traffic management systems has since been demonstrated online in the terminal context \cite{gardi2019multiobjective}. In all of this work the wind is a known input, and the trajectory is re-optimized when the forecast changes, where a published procedure has no such recourse.

\subsection{Delayed Deceleration Approaches}
Dumont \cite{dumont2012fuel} analyzed digital flight data records restricted to $3^\circ$ descent profiles and found a factor-of-two spread in fuel burned below 10{,}000 ft between the most and least efficient flights. Late-decelerating, late-configuring flights burned 30--40\% less fuel. Airspeed profile and time flown with flaps extended were the dominant explanatory variables. Sandberg et al. \cite{sandberg2016delayed} assessed the noise consequences with monitor measurements and modeled contours and found no significant net impact: reduced engine noise during the low-thrust clean segment was offset by added airframe noise at higher airspeed. Thomas and Hansman \cite{thomas2021modeling} developed aircraft-specific deceleration models, predicted 4--8 dB noise reductions below the flight track ahead of the stabilization point, and demonstrated flyability in the 2019 Boeing ecoDemonstrator program, in which a 777-200 flew a DDA with a $3.77^\circ$ final. Their implementation challenges define the present design problem: the procedure must specify where deceleration begins under varying weight and weather, and it must leave adequate margin to the stabilized-approach requirement. Then the remaining attribution question is how much fuel saving comes from delayed deceleration, how much from the steeper final, and how much from the flap schedule.

\subsection{Procedure Design Under Uncertainty}
Ren's  \cite{ren2007modeling} separation methodology is probabilistic in evaluation where TASAT propagates uncertainty in weight, pilot response, FMS logic, and wind through a fixed descent procedure. Kendall and Clarke \cite{kendall2020stochastic} optimized the procedure itself by selecting vertical-navigation waypoint altitudes and airspeeds to minimize expected noise, with TASAT generating the samples. Because each design had roughly a dozen coupled waypoint variables and costly Monte Carlo replications, they used stochastic kriging surrogates with adaptive sampling and differential dynamic programming. The present formulation keeps the same design philosophy, one fixed procedure is judged in expectation over its operating conditions. The objective is fuel, the operational requirement is a chance constraint \cite{charnes1959chance, shapiro2009lectures}, and an optimal-control reduction makes the expectation computable by quadrature without a surrogate model. Section~IV.E gives the computational comparison. The contribution here is attribution: under a common wind climatology and stabilization requirement, the method separates the effects of deceleration architecture, glideslope angle, and flap-deployment schedule.

\section{Approach Procedures and Simulation Environment}

\subsection{Arrival Corridor and Study Aircraft}
\begin{figure}[hbt!]
\centering
\includegraphics[width=\textwidth]{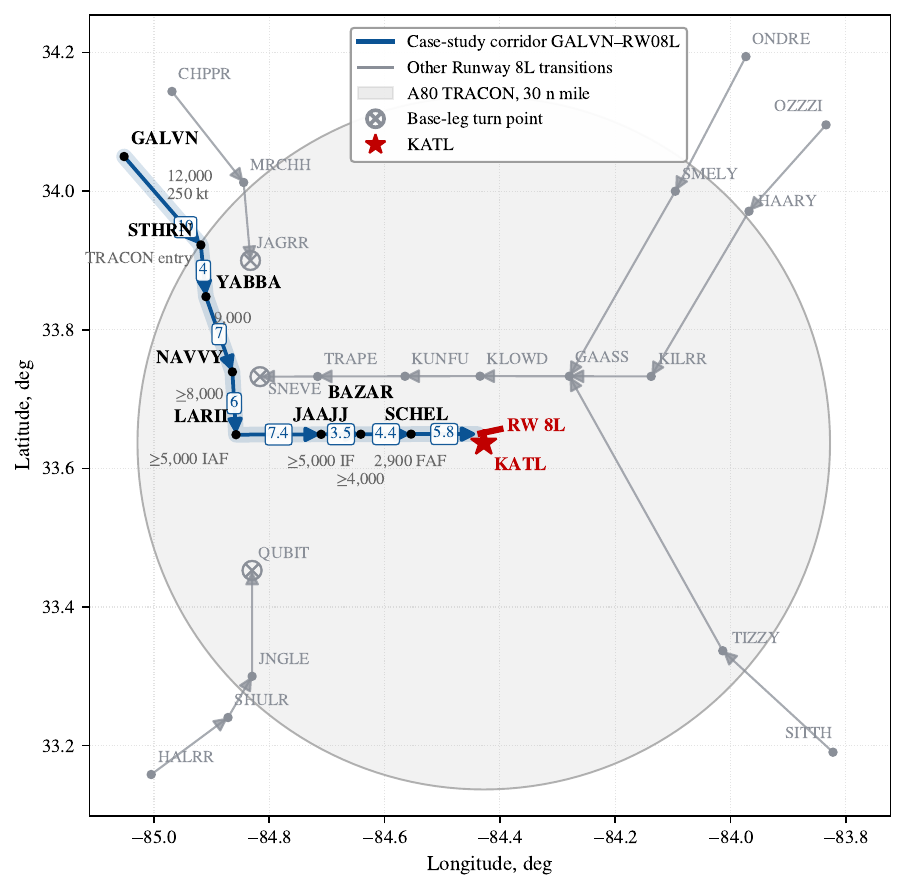}
\caption{Arrival transitions to Runway 8L at KATL in east-flow operations. The case-study corridor (blue) runs from GALVN to the runway threshold, and the gray transitions appear for context.}
\label{fig:corridor}
\end{figure}

All experiments use the straight-in arrival corridor shown in Fig.~\ref{fig:corridor}: the northwest transition of the area-navigation arrivals to Runway 8L in east-flow operations at Hartsfield--Jackson Atlanta International Airport (KATL), field elevation 1{,}026 ft, inside the Atlanta A80 TRACON. The corridor enters at the fix GALVN, 48.1 nautical mile from the threshold at 12{,}000 ft, crosses STHRN at the TRACON boundary and YABBA at 9{,}000 ft, and then descends through a sequence of at-or-above altitude floors (8{,}000 ft at NAVVY, 5{,}000 ft at the initial approach fix LARII and the intermediate fix JAAJJ, and 4{,}000 ft at BAZAR) to the final approach fix (FAF) SCHEL at 5.8 nautical mile. The remaining Runway 8L transitions join the same final approach course through base-leg turns and appear in Fig.~\ref{fig:corridor} for context only. In Fig.~\ref{fig:corridor}, leg lengths are given in nautical mile and altitudes in ft, and the published at-or-above floors carry the $\geq$ mark. The corridor is studied in isolation, consistent with the single-aircraft scope of the procedure design problem. Fuel is metered from GALVN to the runway threshold. Following the steep-final procedure redesign demonstrated in flight test \cite{thomas2021modeling}, the FAF crossing restriction is treated as an at-or-above floor rather than a hard crossing altitude, since a steeper final necessarily crosses the FAF above the $3^\circ$ glideslope crossing altitude. Every candidate design in this study is required to respect all published floors, and designs that do not are discarded as infeasible.

Four airframes are studied, two narrow-body types (A319, B737-800) and two wide-body types (B767-400, A340-300), chosen to span deceleration capability. Table~\ref{tab:aircraft} lists their landing masses, reference speeds, and clean descent speeds. Each aircraft decelerates through five flap detents to its landing configuration with the landing gear extended. Detailed flight histories are shown for one aircraft of each class, the B737-800 and the A340-300.

\begin{table}[hbt!]
\caption{\label{tab:aircraft} Aircraft types studied in this work.}
\centering
\begin{tabular}{lcccc}
\hline
Aircraft & Landing mass, lb & $V_{\mathrm{REF}}$, kt & Clean descent CAS, kt & Landing configuration \\
\hline
A319 & 130{,}000 & 125 & 240 & CONF FULL + gear \\
B737-800 & 146{,}000 & 141 & 240 & Flaps 30 + gear \\
B767-400 & 320{,}000 & 147 & 240 & Flaps 25 + gear \\
A340-300 & 400{,}000 & 122 & 240 & CONF FULL + gear \\
\hline
\end{tabular}
\end{table}

\subsection{The Procedure Architectures: CDA, DDA, and CDDA}
All three procedures in this study descend at idle thrust from the same corridor entry and differ in where the aircraft decelerates and configures.

\emph{Continuous descent approach.} The CDA constrains the altitude and thrust profile. The aircraft descends from the top of descent at or near idle thrust with no low-altitude level-flight segment, following the procedure family developed, flight-tested, charted, and later characterized by optimal control in \cite{clarke1997systems, clarke2004continuous, clarke2013optimized, park2015optimal, park2016vertical}. The speed profile remains conventional. The flap ladder is absorbed during descent: each deceleration segment is flown at the shallow flight-path angle that maintains the minimum descent rate of 500 ft/min \cite{park2015optimal}. The aircraft therefore bleeds speed while descending, extends flaps detent by detent, completes landing configuration before glideslope capture, and flies the final at approach speed \cite{clarke2004continuous}.

\emph{Delayed deceleration approach.} The DDA constrains the speed and configuration profile. The aircraft remains clean at low thrust deep into the arrival and delays deceleration toward final approach speed until the stabilization gate \cite{dumont2012fuel, sandberg2016delayed, thomas2021modeling, fsf2009stabilized}. In the form modeled and flight-demonstrated by Thomas and Hansman \cite{thomas2021modeling}, the late deceleration is flown on a level segment before glideslope capture. That segment is sized so the aircraft can decelerate at idle to the deployment speed of its last approach flap before intercepting the steep final partially configured with gear down. This study reserves DDA for that level-deceleration form. The $3.77^\circ$ final is the ecoDemonstrator value and the maximum permitted by United States criteria for approach category C aircraft; the category D maximum is $3.50^\circ$ \cite{faa_terps}. Thus, $3.77^\circ$ entries for the B737-800 and B767-400 assume relaxed current criteria. International criteria are tighter, with a standard precision-approach maximum of $3.5^\circ$ \cite{icao_pansops}. These criteria govern publishability, not aircraft capability. The delayed-deceleration arms use $3.77^\circ$ for all four airframes, and Sec.~V.B repeats the optimization at $3.50^\circ$ and $3.00^\circ$ to separate architecture from final angle. In both delayed-deceleration arms, landing configuration is completed on final approach. Each remaining flap detent deploys as soon as it is placard-legal after capture, and glideslope capture occurs at the deployment speed of the landing flap.

\emph{Continuous-descent delayed deceleration approach.} The CDDA combines the two constraints. It flies DDA-style late deceleration on the CDA continuous-descent geometry, absorbing the flap ladder on shallow minimum-descent-rate segments with no level flight before capture. Decelerating in descent keeps the aircraft higher for longer at idle and crosses published altitude floors higher, opening capture geometries unavailable to a level deceleration. Section~V.D quantifies the trade. The removed level segment otherwise provides a tailwind-stabilization buffer. The CDA and CDDA share the continuous-descent geometry, idle thrust, and minimum-descent-rate deceleration arcs. They differ in where deceleration must be complete. The CDA reaches approach speed fully configured at glideslope capture, and the CDDA captures at the landing-flap deployment speed and completes deceleration and configuration between capture and the 1{,}000 ft stabilization gate.

The capture altitude and capture distance are geometrically linked through the glideslope angle,
\begin{equation}
\label{eq:hlevel}
h_{\mathrm{level}} = h_{\mathrm{rwy}} + d_{\mathrm{cap}} \tan\gamma_f ,
\end{equation}
so specifying the capture distance $d_{\mathrm{cap}}$ specifies the capture altitude. The ILS glideslope has a standard service volume of 10 nautical mile \cite{faa_aim}, and capture beyond that distance requires an expanded-service-volume authorization and is flagged accordingly in the results.

\subsection{Fast-Time Flight Simulation}
Fuel and stabilization outcomes are computed with a fast-time flight simulation descended from the Tool for the Analysis of Separation and Throughput (TASAT) \cite{ren2007modeling, kendall2020stochastic}, which integrates the rigid-body equations of motion in all six degrees of freedom under closed-loop control. The FMS vertical guidance modes (idle-path descent, altitude and speed hold, glideslope capture and tracking), the pitch autopilot and autothrottle that track them, and the pilot's flap and gear procedures. The state vector, inertial position, attitude, body-axis velocities, controller integrators, and mass, is advanced by a fourth-order Runge-Kutta scheme at a 0.25 s time step, aerodynamic and engine models are type-specific, and fuel flow is integrated along the trajectory. The load-factor and descent-rate histories reported in Sec.~V are direct outputs of this model rather than post-processed point-mass quantities. Flap deployment in the simulation is commanded by a per-detent trigger speed with a distance backstop. The pilot extends the next detent when the airspeed decays to the trigger speed, or upon reaching a backstop distance if the speed trigger has not yet fired, with all backstops clamped at the FAF so that configuration cannot legally be deferred past the stabilization gate. The vertical plan given to the FMS is constructed by integrating the point-mass idle deceleration and descent dynamics backward from the runway threshold, exactly as an FMS builds its descent path, so every candidate design is flyable by construction and the simulation measures what the wind does to it.

\subsection{Wind Model}
The simulation environment inherits the TASAT wind model \cite{ren2007modeling} where the mean-wind profiles varying with altitude, along-track distance, or time. Wind direction changes through descent with a power-law surface taper, Dryden turbulence, and a descent-forecast object for the FMS. Ren \cite{ren2007modeling} used this model for wind variation between successive flights in separation analysis, and Kendall and Clarke \cite{kendall2020stochastic} treated gusts and pilot response delays as process noise along the trajectory. The present procedure-design study uses a reduced wind model tailored to the attribution problem.

\begin{figure}[hbt!]
\centering
\includegraphics[width=.72\textwidth]{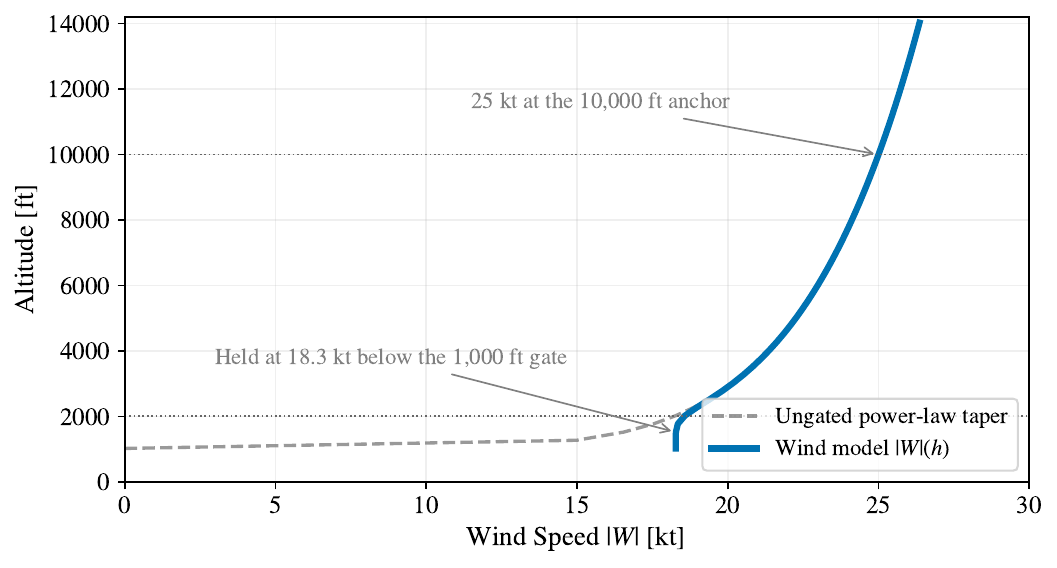}
\caption{Wind realization of Eq.~\eqref{eq:wind} at the truncation magnitude $|w| = 25$ kt. The gate-level value is carried through the final approach segment.}
\label{fig:windmodel}
\end{figure}

The main day-to-day uncertainty on a straight-in corridor is the mean along-corridor wind. Gust-scale turbulence fluctuates about that mean over correlation lengths of hundreds to a few thousand feet, short relative to the 48 nautical mile arrival, so its contribution tends to average out of integrated fuel. Crosswind enters the longitudinal dynamics only through the crab-angle correction to ground speed, a second-order effect on a straight-in track \cite{park2016vertical}. The model therefore represents wind uncertainty by one scalar $w$. The along-corridor wind component at the 10{,}000 ft anchor altitude, positive for tailwind, normally distributed with zero mean and 10 kt standard deviation and truncated at $\pm 25$ kt. Each realization is a height-varying mean-wind table, sampled every 250 ft from the surface to 14{,}000 ft, following the one-seventh-power boundary-layer profile down to the 1{,}000 ft stabilization gate and held constant below it,
\begin{equation}
\label{eq:wind}
\begin{aligned}
W(h; w)
&= w \left(
\frac{\max\{h,\, h_{\mathrm{rwy}} + h_g\} - h_{\mathrm{rwy}}}
{h_a - h_{\mathrm{rwy}}}
\right)^{1/7},\\
h_g
&= 1{,}000~\mathrm{ft}.
\end{aligned}
\end{equation}
at $h_a = 10{,}000$ ft, with the direction held along the corridor at every altitude. Figure~\ref{fig:windmodel} shows the anchor, the boundary-layer taper, and the gate-level freeze at the truncation magnitude $|w| = 25$ kt. Holding the gate-level wind through the final avoids the artificial near-surface shear that would result from tapering the wind to zero at the threshold. Turbulence is disabled, making the simulation deterministic once $(\bm{x}, w)$ is fixed. Enabling the Dryden filter would require Monte Carlo replications, introduce estimator noise, and preclude the exact quadrature of Sec.~IV.B for an effect that is secondary for the fuel objective.

The FMS receives no descent forecast. The vertical plan is constructed at zero wind, and the realized wind affects the aircraft only in flight. This convention represents a nominal published procedure flown in the day's actual wind. Equivalently, $w$ is the forecast error about a zero-mean forecast. Any usable wind knowledge, such as an uplinked descent forecast, would reduce the dispersion the procedure must absorb. The zero-wind plan also keeps every candidate judged on the same nominal path rather than on FMS wind-blending behavior.

The scalar model retains the physics that drives the design problem. Headwinds help deceleration and increase fuel consumption, and tailwinds do the reverse and are where stabilization becomes binding. What the reduction omits, crosswind, gust-scale turbulence, day-to-day variation of the shear exponent, and forecast error about a nonzero forecast, is catalogued with the other limitations in Sec.~V.E.

\section{Stochastic Descent Procedure Optimization}

\subsection{Reduction to a Finite-Dimensional Design Space}
The deterministic layer of the problem is inherited from optimal control. Park and Clarke \cite{park2015optimal} posed the minimum-fuel (and minimum-time) CDA as a multiphase Bolza problem over the point-mass longitudinal dynamics,
\begin{equation}
\label{eq:bolza}
\min_{u(\cdot)} \ J = \sum_{p=1}^{N} \left[ \Phi^{(p)}\big(\bm{z}^{(p)}\big) + \int_{t_0^{(p)}}^{t_f^{(p)}} L^{(p)}\big(\bm{z}^{(p)}, u^{(p)}, t\big)\, \mathrm{d}t \right],
\end{equation}
where $\bm{z}$ and $u$ are the point-mass state (true airspeed, along-track distance, altitude) and control, $N$ is the number of phases, and $\Phi^{(p)}$ and $L^{(p)}$ are the Mayer and Lagrange costs of phase $p$, subject to the dynamic, event, path, and phase-link constraints of the arrival, with the phase boundaries set by the flap-extension speeds and the 250 kt/10{,}000 ft speed-limit interior point. Two of their results carry directly into the present work. First, the optimal solution is dominated by constraint boundary arcs, idle thrust throughout the descent with the airspeed riding the operational and placard limits, and the residual control authority reduces to where the deceleration segments are placed. Second, suboptimal profiles assembled from the vertical-navigation arc family that a certified FMS can actually fly, idle-thrust constant-CAS descent, constant-rate-of-descent deceleration, and constant-flight-path-angle segments, achieve performance nearly indistinguishable from the true optimum, a conclusion that persists in the presence of altitude-dependent wind, where the optimal trajectory is generated by backward and forward integration of the same dynamics \cite{park2016vertical}.

We therefore do not re-solve Eq.~\eqref{eq:bolza} for every candidate procedure. Instead, the search is restricted to that near-optimal, FMS-flyable arc family, built on the longitudinal point-mass dynamics,
\begin{align}
\label{eq:pointmass}
\dot V_T
&= \frac{T - D(h, V_T, \delta)}{m} - g \sin\gamma, \\
\dot h
&= V_T \sin\gamma, \\
\dot d
&= -\big( V_T \cos\gamma + W(h; w) \big).
\end{align}
where $V_T$ is the true airspeed, $\gamma$ the flight-path angle, $\delta$ the flap and gear configuration on which the drag $D$ depends, $d$ the along-track distance to the threshold, and the thrust $T$ is held at its flight-idle value throughout the descent, with the autothrottle recapturing the reference speed only on final approach. Every candidate plan is constructed by integrating Eq.~\eqref{eq:pointmass} backward from the threshold at zero wind ($W \equiv 0$), exactly as in \cite{park2016vertical}, with the flight-path angle prescribed arc by arc. $\gamma$ solved from the constant-CAS condition on the clean idle descent, $\dot h$ fixed at the minimum descent rate of 500 ft/min on the deceleration segments, and $\gamma = -\gamma_f$ on the final. The remaining freedom is the finite-dimensional vector,
\begin{equation}
\label{eq:designset}
\bm{x} = \big(d_{\mathrm{cap}},\, x_1, \ldots, x_G\big) \in \mathcal{X} = \mathcal{D} \times \prod_{g=1}^{G} [-\eta_g, +\eta_g],
\end{equation}
where $\mathcal{D}$ is the admissible set of capture distances and $G \le 5$ is the number of flap-detent groups. Because the optimal-control analysis fixes the arc structure, the infinite-dimensional trajectory problem reduces to a low-dimensional parameter optimization. This reduction makes the exact treatment of uncertainty in Sec.~IV.B computationally practical.

For the delayed-deceleration arms, the ILS glideslope capture distance $\mathcal{D}$ spans 6 to 10 nautical mile in 0.5 nautical mile steps. The lower bound keeps capture outside the FAF, and the upper bound is the edge of the standard glideslope service volume. For the $3^\circ$ CDA, the published 5{,}000 ft altitude floors make capture distances much below 12 nautical mile geometrically infeasible. The CDA grid therefore extends to 13 nautical mile as a published-procedure exception. This asymmetry is discussed in Sec.~V. The glide-path variants of Sec.~V.B use the same construction, with grid limits adjusted for final angle as listed in Sec.~IV.F.

Each flap detent $i$ has a placard window $[V_{\min,i}, V_{\max,i}]$ within which deployment is acceptable (i.e., the flap triggering speed). Consecutive detents with identical windows are collected into groups $g = 1, \ldots, G$, and each group receives a single offset $x_g$ about its window midpoint,
\begin{equation}
\label{eq:trigger}
\begin{aligned}
\bar V_g
&= \mathrm{round}\!\left[
\min\!\big\{ \max( \mu_g + x_g,\ V_{\min,g}),\ V_{\max,g} \big\}
\right],\\
x_g
&\in [-\eta_g, +\eta_g].
\end{aligned}
\end{equation}
where $\mu_g$ and $\eta_g$ are the window midpoint and half-width and the rounding quantizes to 1 kt. The realized per-detent triggers are made monotone by the running minimum
\begin{equation}
\label{eq:cascade}
\begin{aligned}
V_i
&= \min_{j \le i} \bar V_{g(j)},~for ~i=1, \ldots, 5.
\end{aligned}
\end{equation}
so a later detent cannot be assigned a faster trigger than an earlier one. Quantization and the cascade of Eq.~\eqref{eq:cascade} make the offset-to-trigger map many-to-one, collapsing $\mathcal{X}$ onto a finite lattice of a few hundred distinct designs per airframe. The search in Sec.~IV.D exploits this lattice. Positive offsets deploy flaps early, adding drag when deceleration needs help; negative offsets deploy late, keeping the aircraft clean longer. The placard window is the legal envelope. Park and Clarke note that extension speed may be chosen anywhere within it \cite{park2015optimal}, and flight-data analysis links approach fuel burn to airspeed and time with flaps extended \cite{dumont2012fuel}. Trigger speeds control both quantities, so Eqs.~\eqref{eq:trigger}--\eqref{eq:cascade} make them decision variables.

In flight, the triggers act through hitting times. Detent $i$ deploys at,
\begin{equation}
\label{eq:flaplaw}
\begin{aligned}
\tau_i
&= \inf\Big\{ t :
\big( V_{\mathrm{CAS}}(t) \le V_i
\ \text{or}\ d(t) \le d_i^{\mathrm{bs}} \big),
\ V_{\mathrm{CAS}}(t) \le V_{\max,i} \Big\},\\
d_i^{\mathrm{bs}}
&= \max\big\{ d_i^{\mathrm{plan}},\, d_{\mathrm{FAF}} \big\}.
\end{aligned}
\end{equation}
where $V_{\mathrm{CAS}}$ is calibrated airspeed, $d_i^{\mathrm{plan}}$ is the distance at which the zero-wind plan crosses $V_i$, and $d_{\mathrm{FAF}}$ is the final approach fix distance. The speed condition carries the design variable. The distance backstop fires when a tailwind keeps airspeed above the trigger beyond the planned deployment point, and the placard condition defers deployment until extension is legal. The clamp at $d_{\mathrm{FAF}}$ prevents deliberate deferral past the stabilization gate and any remaining configuration failure reflects a physical deceleration deficit.

Capture distance and trigger offsets are the only optimized quantities. Top of descent is an output of plan construction. In the cruise-phase problem, top of descent is the natural decision \cite{park2015optimal} because an extendable cruise segment absorbs idle-descent slack. The terminal-area problem is constrained from the other end. The entry state, published crossing altitudes, and final angle determine the profile, leaving the primary freedom near the bottom where deceleration occurs. Once $\bm{x}$ is chosen, backward integration of Eq.~\eqref{eq:pointmass} places the top of descent where the idle path meets the published crossing altitudes, with any excess distance flown level. Searching over $d_{\mathrm{cap}}$ therefore searches over top-of-descent placement in a coordinate bounded by the FAF and glideslope service volume. An independently chosen top of descent would overdetermine the profile and break the idle-thrust arcs on which the near-optimality of the arc family rests.

\subsection{Chance-Constrained Stochastic Program}
Let the wind parameter $W$ be a random variable with the truncated-normal law $\mathrm{TN}(0, \sigma_w^2;\, [w_{\min}, w_{\max}])$, $\sigma_w = 10$ kt on $[-25, +25]$ kt, whose density is
\begin{equation}
\label{eq:density}
\begin{aligned}
\varphi_{\mathrm{TN}}(w)
&= \frac{\exp\!\big( -w^2 / 2\sigma_w^2 \big)}
{\displaystyle\int_{w_{\min}}^{w_{\max}}
\exp\!\big( -v^2 / 2\sigma_w^2 \big)\, \mathrm{d}v}, ~~ w\in [w_{\min}, w_{\max}].
\end{aligned}
\end{equation}
each realization acting on the aircraft through the shear profile of Eq.~\eqref{eq:wind}. For a design $\bm{x} \in \mathcal{X}$, the simulation of Sec.~III.C delivers the fuel functional
\begin{equation}
\label{eq:fuel}
f(\bm{x}, w) = \int_{t_{\mathrm{entry}}}^{t_{\mathrm{thr}}} \dot m_f(t)\, \mathrm{d}t,
\end{equation}
the integral of the type-specific fuel-flow model along the closed-loop trajectory from the corridor-entry crossing $t_{\mathrm{entry}}$ to the threshold crossing $t_{\mathrm{thr}}$, together with the stabilized-approach indicator $s(\bm{x}, w) \in \{0, 1\}$ constructed in Sec.~IV.C. Both are deterministic maps once $(\bm{x}, w)$ is fixed. The design problem is the chance-constrained stochastic program \cite{charnes1959chance, shapiro2009lectures}
\begin{align}
\label{eq:exactproblem}
\min_{\bm{x} \in \mathcal{X}} \quad
& \mathbb{E}\big[ f(\bm{x}, W) \big]
= \int_{w_{\min}}^{w_{\max}}
f(\bm{x}, w)\, \varphi_{\mathrm{TN}}(w)\, \mathrm{d}w \\
\text{subject to} \quad
& \mathbb{P}\big[ s(\bm{x}, W) = 1 \big]
= \int_{w_{\min}}^{w_{\max}}
s(\bm{x}, w)\, \varphi_{\mathrm{TN}}(w)\, \mathrm{d}w
\ge 1 - \varepsilon. \nonumber
\end{align}
with $\varepsilon = 0.05$. Minimize the climatological-average fuel over designs that deliver a stabilized approach on at least 95\% of arrivals.

Because $W$ is scalar with a known density, the expectation and the probability in Eq.~\eqref{eq:exactproblem} are one-dimensional integrals, and we discretize them by deterministic quadrature rather than Monte Carlo sampling. On $K$ equispaced nodes $w_k = w_{\min} + (k-1)\, \Delta w$ with normalized weights,
\begin{equation}
\label{eq:weights}
\begin{aligned}
\omega_k
&= \frac{\exp\!\big( -w_k^2 / 2\sigma_w^2 \big)}
{\sum_{j=1}^{K} \exp\!\big( -w_j^2 / 2\sigma_w^2 \big)}, ~k
= 1, \ldots, K.
\end{aligned}
\end{equation}
the discretized design problem, written out in full with the trigger construction, the capture geometry, the published-procedure feasibility requirement, and the box constraints of the design set, is
\begin{subequations}
\label{eq:problem}
\begin{align}
\min_{\bm{x} = (d_{\mathrm{cap}},\, x_1, \ldots, x_G)} \quad
& \sum_{k=1}^{K} \omega_k\, f(\bm{x}, w_k) \label{eq:probobj} \\
\text{subject to} \quad
& \sum_{k=1}^{K} \omega_k\, s(\bm{x}, w_k) \ \ge\ 1 - \varepsilon, \label{eq:probchance} \\
& V_i = \min_{j \le i} \bar V_{g(j)},
&& i = 1, \ldots, 5,
\quad \bar V_g \ \text{from Eq.~\eqref{eq:trigger}}, \label{eq:probtrigger} \\
& h_{\mathrm{level}} = h_{\mathrm{rwy}} + d_{\mathrm{cap}} \tan\gamma_f, \label{eq:probgeom} \\
& h^{\mathrm{plan}}(d_r;\, \bm{x}) \ge h_r^{\mathrm{floor}},
&& r = 1, \ldots, R, \label{eq:probfloors} \\
& d_{\mathrm{cap}} \in \mathcal{D}, \qquad x_g \in [-\eta_g, +\eta_g],
&& g = 1, \ldots, G. \label{eq:probbox}
\end{align}
\end{subequations}
where $f(\bm{x}, w_k)$ and $s(\bm{x}, w_k)$ are generated by the simulation of Sec.~III.C under the planning dynamics of Eq.~\eqref{eq:pointmass}, the deployment law of Eq.~\eqref{eq:flaplaw}, and the indicator of Eq.~\eqref{eq:indicator} below. Each node is one deterministic simulation: the scalar $w_k$ is materialized as the steady shear profile of Eq.~\eqref{eq:wind} and the aircraft flies the zero-wind plan through it, the closed-loop guidance, autothrottle, and flap logic absorbing the difference, so the wind enters the flown trajectory but never the plan. In Eq.~\eqref{eq:probfloors}, $h^{\mathrm{plan}}(d_r; \bm{x})$ is the altitude of the zero-wind backward plan at the along-track distance $d_r$ of the $r$th published crossing restriction and $h_r^{\mathrm{floor}}$ the corresponding at-or-above floor (Sec.~III.A). A design whose plan descends below any floor is discarded as infeasible before any wind is simulated. Equation~\eqref{eq:problem} is a sample-average approximation of Eq.~\eqref{eq:exactproblem} \cite{shapiro2009lectures} on a fixed, pdf-weighted scenario set: every candidate design is evaluated against the identical wind ensemble (the common-random-numbers property holds by construction), the objective is exactly reproducible with no estimator variance, and the only approximation error is quadrature error, controlled as described in Sec.~IV.D. The design grid uses $\Delta w = 5$ kt ($K = 11$) and verification uses $\Delta w = 1$ kt ($K = 51$). Both grids discretize the same continuous distribution where the node spacing is a numerical integration mesh, not a wind-forecast or measurement resolution, and no per-flight wind information enters the design, which is the premise of a published procedure flown open-loop against the climatology.

Read line by line, Eq.~\eqref{eq:problem} makes the design problem explicit. The decision variables are the glideslope-capture distance $d_{\mathrm{cap}}$, which through Eq.~\eqref{eq:probgeom} sets the altitude at which the shallow descent hands over to the $\gamma_f$ final and therefore how long the aircraft stays fast and clean, and the flap trigger offsets $x_1, \ldots, x_G$, one per placard-window group (Eqs.~\eqref{eq:trigger}--\eqref{eq:cascade}), which schedule the drag that flight-data analysis identifies as the dominant correlate of approach fuel \cite{dumont2012fuel}. The objective (Eq.~\eqref{eq:probobj}) is the pdf-weighted expected fuel from corridor entry to threshold. The chance constraint (Eq.~\eqref{eq:probchance}) requires a stabilized approach on at least $100(1-\varepsilon)\% = 95\%$ of arrivals, pricing the rare failures as go-arounds rather than forbidding them. The structural constraints (Eqs.~\eqref{eq:probtrigger}--\eqref{eq:probbox}) enforce the monotone trigger cascade, the capture geometry, and the published crossing floors that discard an infeasible plan before any wind is simulated.

\subsection{Stabilization Chance Constraint}
The indicator in Eq.~\eqref{eq:exactproblem} encodes the standard stabilized-approach criteria \cite{fsf2009stabilized, thomas2021modeling}: $s(\bm{x}, w) = 1$ if and only if the simulated arrival satisfies the following criteria,
\begin{enumerate}
\item Landing configuration established at the 1{,}000 ft-above-field gate,
\item Calibrated airspeed at the gate no greater than $V_{\mathrm{REF}} + 15$ kt,
\item Calibrated airspeed at the threshold no less than $V_{\mathrm{REF}} - 10$ kt,
\item Peak longitudinal deceleration along the approach no greater than 0.12 g.
\end{enumerate}
Formally, let $t_{\mathrm{gate}} = \inf\{ t : h(t) \le h_{\mathrm{rwy}} + 1{,}000\ \mathrm{ft} \}$ and $t_{\mathrm{thr}} = \inf\{ t : d(t) \le 0 \}$ be the gate and threshold crossing times of the simulated arrival, $\delta(t)$ its instantaneous flap and gear configuration with $\delta_{\mathrm{land}}$ the landing configuration, and $n_x(t)$ its longitudinal load factor, negative when decelerating. The indicator is the product of the four event indicators,
\begin{equation}
\label{eq:indicator}
s(\bm{x}, w) =
\mathbb{I}\big\{ \delta(t_{\mathrm{gate}}) = \delta_{\mathrm{land}} \big\}\;
\mathbb{I}\big\{ V_{\mathrm{CAS}}(t_{\mathrm{gate}}) \le V_{\mathrm{REF}} + 15 \big\}\;
\mathbb{I}\big\{ V_{\mathrm{CAS}}(t_{\mathrm{thr}}) \ge V_{\mathrm{REF}} - 10 \big\}\;
\mathbb{I}\big\{ \min_t\, n_x(t) \ge -0.12 \big\},
\end{equation}
evaluated on the same trajectory that generates the fuel integral of Eq.~\eqref{eq:fuel}, so fuel and stabilization are never computed from inconsistent runs. The constraint is probabilistic rather than absolute value. A design may fail to stabilize in a rare tailwind provided the weighted probability of such failures stays within the risk budget. In operation such a case would end in a routine go-around, the chance constraint bounds how often the procedure is allowed to demand one. Designs whose trigger speeds cannot become placard-legal in time under some wind, a deceleration deadlock in which idle thrust on the steep final cannot slow a clean airframe to the next legal deployment speed, simply fail criterion 1 at those nodes and are pruned by the same constraint, so the trigger optimization also determines which designs are feasible at all.

The gate of criteria 1 and 2 sits at 1{,}000 ft above the field rather than at the final approach fix. $V_{\mathrm{REF}}$ is the reference landing speed of the threshold crossing. The speed flown on the stabilized final is the approach speed, $V_{\mathrm{REF}} + 5$ kt plus wind uncertainties, and the industry criteria require the approach to be stabilized (in landing configuration, on the glidepath, no slower than $V_{\mathrm{REF}}$ and no faster than $V_{\mathrm{REF}} + 20$ kt) by 1{,}000 ft above field elevation in instrument conditions \cite{fsf2009stabilized}. No operational requirement pins the airspeed at the FAF itself; terminal speed control presumes the opposite, since controllers may assign arriving turbojets 170 kt or more all the way down to the fix and are barred from issuing speed adjustments only inside it \cite{faa2023atc}, so crossing the FAF tens of knots above the approach speed is routine line practice, and the delayed-deceleration demonstrations define the procedure on that margin, with the deceleration completing at the stabilization point rather than at the fix \cite{thomas2021modeling}. The simulated procedures also reflect this. The guidance targets the approach speed from the FAF inward, and the CDA, configured well upstream, meets it at the fix. The wide-body CDDA, whose glideslope capture sits only a few miles outside the FAF, is still decelerating as it crosses and completes the deceleration between the fix and the gate, where criteria 1 and 2 judge it. Requiring the approach speed at the FAF instead would declare the delayed-deceleration architecture infeasible by construction, contradicting the flight demonstrations reviewed in Sec.~II.

The common industry practice carry one further element, a sink rate no greater than 1{,}000 ft/min on the stabilized final, with the explicit rider that an approach requiring a higher sink rate calls for a special briefing rather than being prohibited \cite{fsf2009stabilized}. On a glideslope the final sink rate is fixed by geometry, the ground speed times the tangent of the final angle, so it is a property of the published angle and the approach speed rather than of the design variables the optimization controls. The element is therefore monitored rather than folded into Eq.~\eqref{eq:indicator}. Every simulation records the sink rate at the 1{,}000 ft gate and its maximum from the gate to the threshold, and any design whose stabilized final exceeds 1{,}000 ft/min at some wind node is flagged as requiring the special briefing. Section~V.C reports these flags. They attach almost exclusively to the steep-final arms and are part of the operational cost of the steeper final.

\subsection{Search and Verification}
The maps from the design $\bm{x}$ to the fuel $f$ and to the indicator $s$ are evaluated by simulation and are discontinuous in the design: a 1 kt trigger change can move a flap deployment across a guidance-mode boundary. Both maps also depend on $\bm{x}$ only through the realized design
\begin{equation}
\label{eq:lattice}
\begin{aligned}
q(\bm{x})
&= \big( d_{\mathrm{cap}},\, V_1, \ldots, V_5 \big),\\
f(\bm{x}, w)
&= \tilde f\big( q(\bm{x}), w \big),\\
s(\bm{x}, w)
&= \tilde s\big( q(\bm{x}), w \big).
\end{aligned}
\end{equation}
so the effective search space is the finite lattice $\Lambda = q(\mathcal{X})$ of a few hundred distinct designs per airframe, and the search is derivative-free with real-time deployment capabilities. Stage 1 sweeps a coarse grid over $d_{\mathrm{cap}}$ and a common normalized offset applied to all flap groups. Stage 2 refines the stage-1 incumbent by block-coordinate descent, optimizing one group offset at a time over a finer offset grid, then the capture distance, and cycling until no improvement. Candidates that map to an already-evaluated point of $\Lambda$ are served from cache.

Stage 3 controls the quadrature error of the constraint. For fixed $\bm{x}$, the fuel $f(\bm{x}, \cdot)$ is piecewise smooth in $w$, so the 5 kt rule integrates it accurately. The indicator behaves differently. Its failure set
\begin{equation}
\label{eq:failset}
\begin{aligned}
\mathcal{F}(\bm{x})
&= \big\{ w \in [w_{\min}, w_{\max}] : s(\bm{x}, w) = 0 \big\}\\
&= \bigcup_{j=1}^{J(\bm{x})} (a_j, b_j).
\end{aligned}
\end{equation}
is a finite union of wind intervals whose endpoints $a_j$ and $b_j$ are not known in advance, and the quadrature of Eq.~\eqref{eq:probchance} resolves $\mathcal{F}(\bm{x})$ only to the width of its cells,
\begin{equation}
\label{eq:quaderror}
\left|
\sum_{k=1}^{K} \omega_k\, s(\bm{x}, w_k)
- \mathbb{P}\big[ s(\bm{x}, W) = 1 \big]
\right|
\le 2\, J(\bm{x}) \max_{1 \le k \le K} \omega_k ,
\end{equation}
since each interval endpoint can be misplaced by at most one cell and each cell carries at most $\max_k \omega_k$ of probability mass. On the 5 kt design grid $\max_k \omega_k \approx 0.20$ at the mode of the wind law, so a narrow tailwind band can escape the coarse grid entirely. A design can stabilize at the $+15$ and $+20$ kt nodes yet fail throughout $(+16, +19)$. On the 1 kt verification grid the same bound falls to about 0.04 per interval endpoint. Every candidate is therefore re-evaluated on the 1 kt grid in order of coarse-grid expected fuel, and the accepted optimum is defined as,
\begin{equation}
\label{eq:accept}
\begin{aligned}
\bm{x}^\star
&= \operatorname*{arg\,min}_{\bm{x} \in \Lambda}
\left\{
\sum_{k} \omega_k^{(5)}\, f\big(\bm{x}, w_k^{(5)}\big)
:
\sum_{k} \omega_k^{(1)}\, s\big(\bm{x}, w_k^{(1)}\big)
\ge 1 - \varepsilon
\right\}.
\end{aligned}
\end{equation}
where the superscripts distinguish the 5 kt and 1 kt grids. The objective is ranked on the design grid, and feasibility is certified on the verification grid. During the design iterations, the promotion step rejected several coarse-grid winners whose stabilization probability fell to 0.944 inside narrow tailwind bands that the coarse grid straddled.

\subsection{Relation to Prior Stochastic Formulations}
Equation~\eqref{eq:exactproblem} has the same abstract form as the noise-abatement program of Kendall and Clarke \cite{kendall2020stochastic}, but the computational setting differs. In their problem, the design is the vertical geometry itself, with roughly a dozen coupled continuous VNAV-waypoint variables, and uncertainty enters as process noise. Each design evaluation is therefore a Monte Carlo estimate, and direct tensor-product evaluation is impractical, stochastic-kriging surrogates and differential dynamic programming are appropriate. Here the optimal-control reduction of Sec.~IV.A fixes the arc structure, leaving a few hundred distinct designs. The uncertainty is a scalar with known density, and the simulation is deterministic given $(\bm{x}, w)$. Equation~\eqref{eq:problem} can therefore be evaluated by one-dimensional quadrature for every candidate, without a surrogate model.

\subsection{Study Setups}
The procedure arms optimized under the identical formulation, wind law, and stabilization constraint differ only in the architecture of the deceleration and the final approach.
\begin{itemize}
\item \emph{CDA}: Decelerates and configures on the shallow minimum-descent-rate segments and intercepts the glideslope fully configured. The published altitude floors push its feasible captures to roughly 12 nautical mile and beyond, so its capture grid extends to 13 nautical mile (Sec.~IV.A).
\item \emph{CDDA}: Holds the clean 240 kt descent speed deeper into the arrival, flies the same continuous-descent deceleration later with no level flight before intercept, captures at the deployment speed of its landing flap, and completes the configuration on final. Its capture grid spans 6--10 nautical mile, from just outside the FAF to the edge of the standard glideslope service volume.
\item \emph{DDA}: Identical to the CDDA except that the deceleration is flown on a level segment at the capture altitude, as in the demonstration campaign \cite{thomas2021modeling}. It enters the robustness comparison of Sec.~V.D.
\item \emph{Glide-path variants}: Both architectures re-optimized with only the final angle changed, at the standard $3.00^\circ$, the approach category D maximum of $3.50^\circ$, and the $3.77^\circ$ category C maximum, each for all four airframes. These arms complete the architecture-angle factorial. The $3.00^\circ$ pair isolates delayed deceleration on the standard final, the $3.50^\circ$ pair uses the steepest final publishable for every study airframe under current criteria, and the $3.77^\circ$ pair tests the edge of those criteria. Because published floors move the shallowest feasible capture with angle through Eq.~\eqref{eq:hlevel}, the $3.00^\circ$ and $3.50^\circ$ capture grids extend past the standard service volume on the same published-procedure exception as the conventional CDA, reaching the 5{,}000 ft platform at 12.48 and 10.69 nautical mile.
\end{itemize}
Each arm carries its own optimized capture distance and trigger offsets. Two fixed trigger ladders anchor the comparison against current practice, evaluated for every architecture and final angle at the corridor's current 5{,}000 ft capture altitude (12.48 nautical mile on the $3.00^\circ$ final, 10.69 on the $3.50^\circ$, and 9.93 on the $3.77^\circ$) and reported for the headline $3.00^\circ$ CDA and $3.50^\circ$ CDDA pair in Table~\ref{tab:main}. The \emph{minimum-speed rule}, the common line practice of extending each next flap detent as speed decays to just above the minimum speed of the current detent (here, that minimum plus 10 kt), and the \emph{midpoint rule}, deploying each flap at the middle of its placard window ($x_g = 0$).

\section{Results and Discussion}

\subsection{Wind-Conditional Performance of the Optimized Designs}
\begin{figure}[hbt!]
\centering
\begin{subfigure}{0.49\textwidth}
\includegraphics[width=\textwidth]{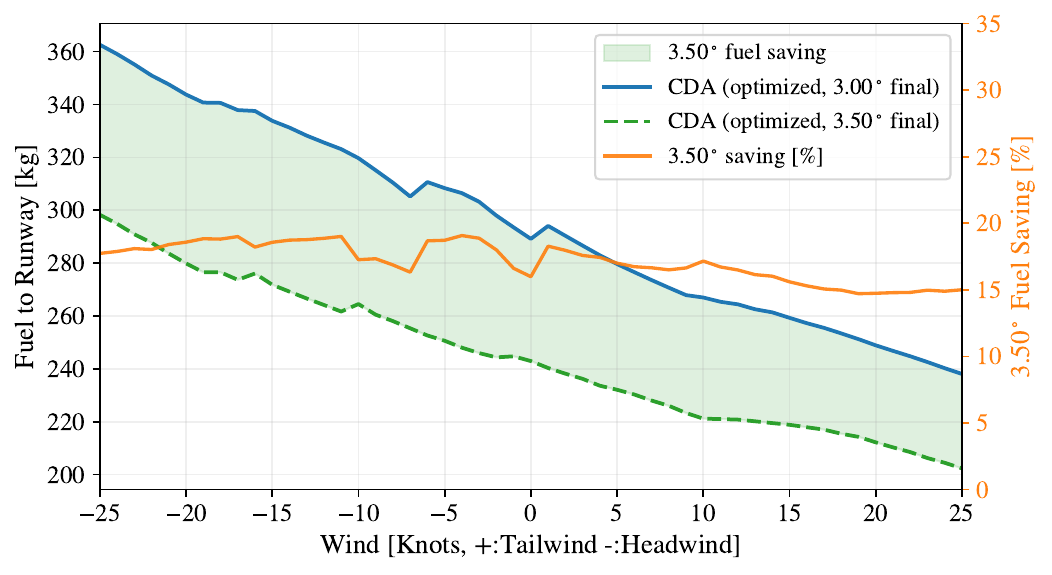}
\caption{B737-800 fuel consumption.}
\end{subfigure}
\begin{subfigure}{0.49\textwidth}
\includegraphics[width=\textwidth]{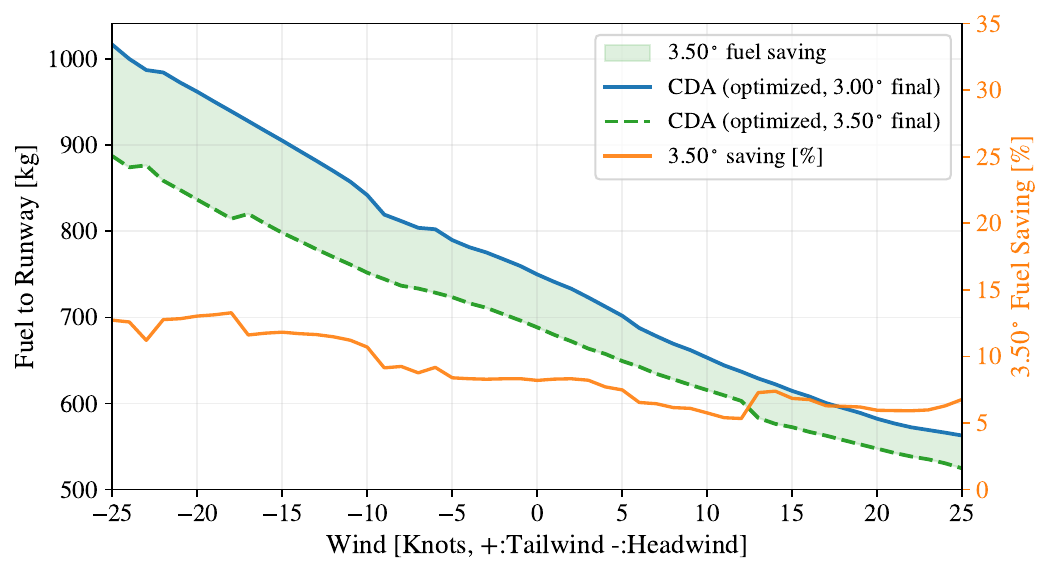}
\caption{A340-300 fuel consumption.}
\end{subfigure}
\\
\begin{subfigure}{0.49\textwidth}
\includegraphics[width=\textwidth]{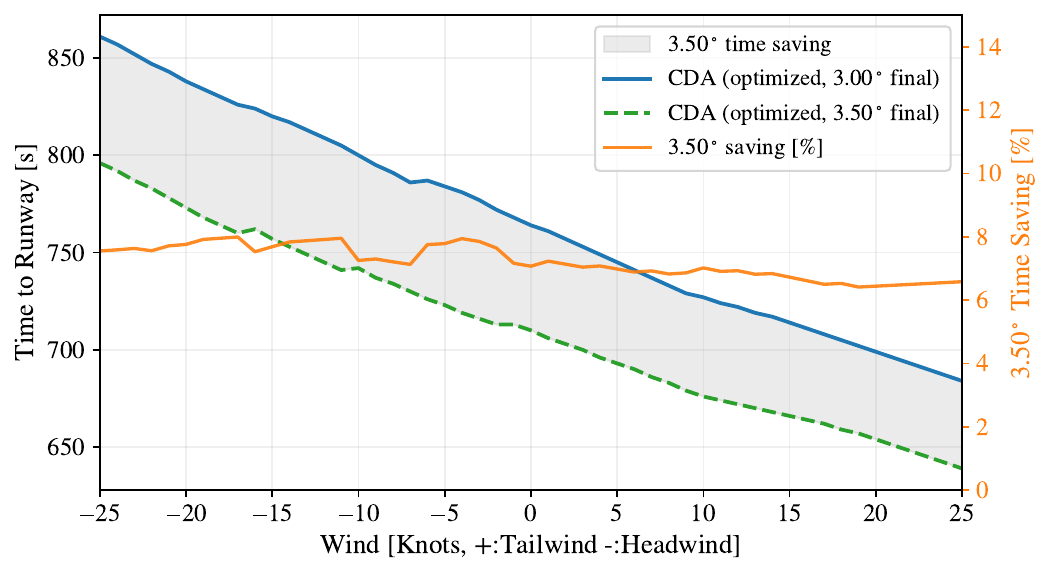}
\caption{B737-800 flight time.}
\end{subfigure}
\begin{subfigure}{0.49\textwidth}
\includegraphics[width=\textwidth]{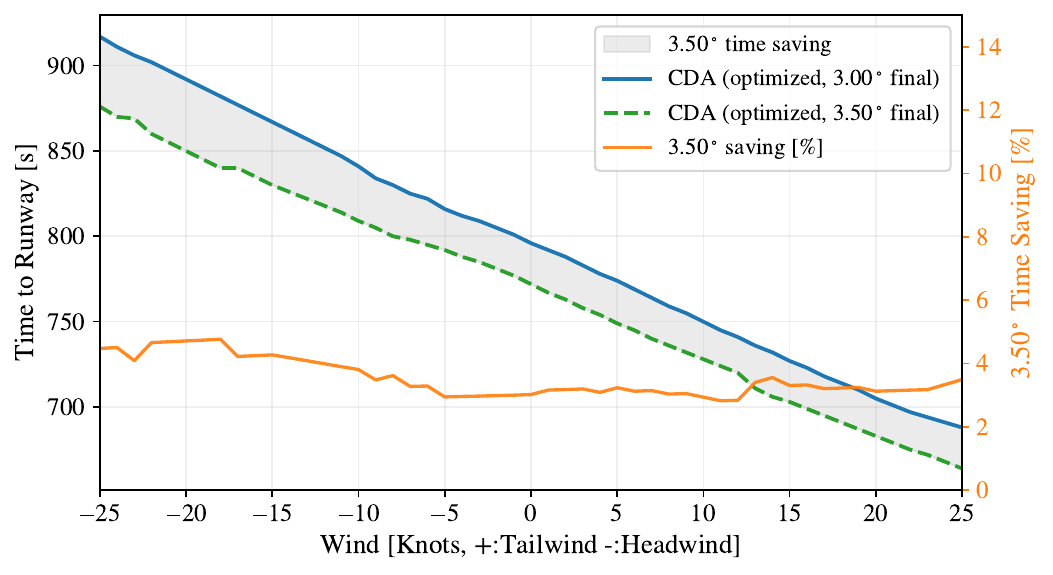}
\caption{A340-300 flight time.}
\end{subfigure}
\caption{Fuel and flight time from the TRACON boundary to the runway threshold for the optimized $3.00^\circ$ and $3.50^\circ$ CDA designs across the wind range.}
\label{fig:sweep}
\end{figure}

Figure~\ref{fig:sweep} shows the main effect in the results. The conventional CDA optimized on the standard $3.00^\circ$ final and re-optimized on the $3.50^\circ$ Category D maximum, evaluated across the full wind range at 1 kt resolution for the two detailed case-study airframes. In each panel of Fig.~\ref{fig:sweep}, the shading spans the two designs at each wind node, and the right-hand axis reports the saving of the steeper final in percent. Holding architecture fixed isolates the benefit of final angle and capture geometry. The $3.50^\circ$ final burns less fuel at every wind node: 14.7--19.1\% less for the B737-800 and 5.3--13.3\% less for the A340-300. It also reaches the runway 45--66 s earlier for the B737-800 and 21--42 s earlier for the A340-300, because the shorter capture stretches the clean idle descent. The fuel curves are not strictly monotone in wind. Each design flies a fixed zero-wind plan, and flap deployment and thrust spool-up are threshold events. A 1 kt wind change can move one of these events by up to about half a nautical mile and produce the small excursions visible in the curves. The matched-glideslope architecture effect is smaller, with 3--8\% at $3.00^\circ$ and 0.3--3.9\% at $3.50^\circ$. Section~V.B gives the full factorial, and Sec.~V.C reports the headline procedures and baselines.

\subsection{Deceleration Architecture versus Glideslope Angle}
\begin{figure}[hbt!]
\centering
\includegraphics[width=\textwidth]{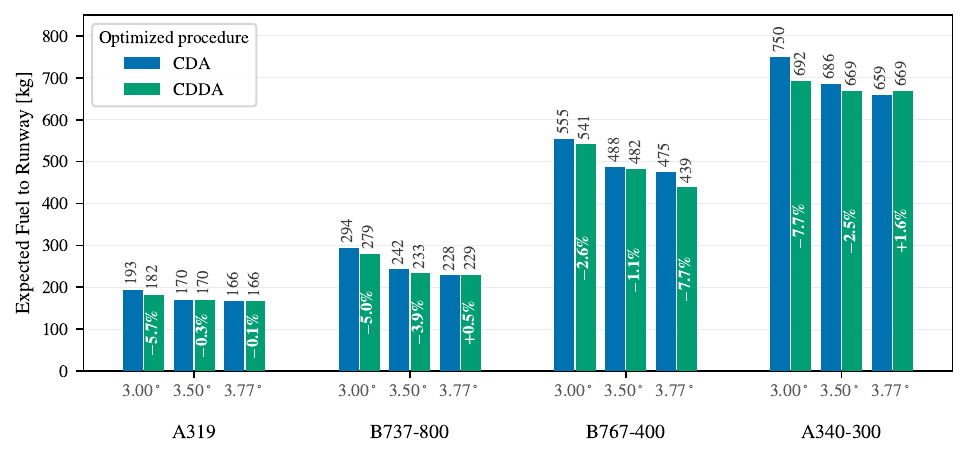}
\caption{Expected fuel of the optimized CDA and the optimized CDDA paired at each final angle for the designs of Table~\ref{tab:ablation}.}
\label{fig:ablation}
\end{figure}

The demonstrated DDA differs from the conventional CDA in delayed deceleration and final angle. Section~III.B defined the regulatory boundary. $3.77^\circ$ is the category C procedure-design maximum, $3.50^\circ$ is the category D maximum for the B737-800 and B767-400, and the $3.77^\circ$ entries for those two airframes assume relaxed current criteria. Table~\ref{tab:ablation} and Fig.~\ref{fig:ablation} separate architecture and angle by optimizing both procedures under the same formulation at $3.00^\circ$, $3.50^\circ$, and $3.77^\circ$ for all four airframes. Each saving in this subsection is therefore a matched-glideslope comparison. In Fig.~\ref{fig:ablation}, the bar labels give expected fuel in kilograms, and the in-bar label on each CDDA bar is the signed change relative to the optimized CDA at the same angle, negative when the optimized CDA burns less. The $3.00^\circ$ CDA and $3.50^\circ$ CDDA designs reappear in Table~\ref{tab:main}, which reports their trigger ladders and fixed-rule baselines.

\begin{table}[hbt!]
\caption{\label{tab:ablation} The same-angle factorial with both architectures optimized at each final angle under the identical formulation ($3.00^\circ$ standard, $3.50^\circ$ category D maximum, $3.77^\circ$ category C maximum). The saving on each CDDA row is referenced to the optimized CDA at the same final angle, and negative values mean the optimized CDA burns less.}
\centering
\small
\setlength{\tabcolsep}{3.5pt}
\begin{tabular}{lcccccccc}
\hline
 &  & \multicolumn{3}{c}{CDA optimized} & \multicolumn{4}{c}{CDDA optimized} \\
\cline{3-5}\cline{6-9}
Aircraft & Final, deg & $d_{\mathrm{cap}}$ & $E[f]$ & $P_s$ & $d_{\mathrm{cap}}$ & $E[f]$ & Saving, \% & $P_s$ \\
\hline
A319 & 3.00 & 11.5 & 193.3 & 0.985 & 12.0 & 182.3 & 5.7 & 0.955 \\
 & 3.50 & 9.0 & 170.2 & 0.965 & 9.5 & 169.7 & 0.3 & 0.955 \\
 & 3.77 & 8.0 & 166.3 & 0.965 & 7.0 & 166.1 & 0.1 & 0.955 \\
\hline
B737-800 & 3.00 & 11.5 & 293.8 & 1.000 & 11.5 & 279.2 & 5.0 & 1.000 \\
 & 3.50 & 7.0 & 242.5 & 1.000 & 7.0 & 233.0 & 3.9 & 0.980 \\
 & 3.77 & 6.0 & 228.0 & 0.955 & 6.5 & 229.1 & $-0.5$ & 0.955 \\
\hline
B767-400 & 3.00 & 11.0 & 555.1 & 1.000 & 11.5 & 540.8 & 2.6 & 1.000 \\
 & 3.50 & 8.5 & 488.0 & 1.000 & 9.0 & 482.5 & 1.1 & 1.000 \\
 & 3.77 & 8.0 & 475.1 & 1.000 & 7.5 & 438.6 & 7.7 & 0.996 \\
\hline
A340-300 & 3.00 & 11.5 & 750.4 & 1.000 & 11.5 & 692.3 & 7.7 & 0.965 \\
 & 3.50 & 9.0 & 686.3 & 0.955 & 9.0 & 669.0 & 2.5 & 0.973 \\
 & 3.77 & 8.0 & 658.8 & 0.955 & 9.5 & 669.4 & $-1.6$ & 0.955 \\
\hline
\end{tabular}
\end{table}

Delayed deceleration alone leads to 2.6--7.7\% of fuel saving. Re-optimized on the standard $3.00^\circ$ final, the CDDA saves 5.7\% (A319), 5.0\% (B737-800), 2.6\% (B767-400), and 7.7\% (A340-300) relative to the optimized CDA. The capture distances explain the modest size. On the shared $3.00^\circ$ final, the published floors bind both procedures, and every $3.00^\circ$ CDDA optimum captures at 11.5--12.0 nautical mile, beyond the standard service volume and close to the CDA geometry. The remaining difference is architecture, where the CDDA captures at landing-flap deployment speed and completes configuration on final, delaying deceleration by a few nautical mile on the shallow glideslope.

The steeper final closes most of the architecture gap. A steeper final crosses published floors closer to the runway, moves feasible capture inside the service volume, stretches the idle descent, and shortens the configured final segment. These benefits do not depend on delayed deceleration. Re-optimized on the $3.77^\circ$ final, the CDA captures at 6.0--8.0 nautical mile and burns 12--22\% less than its $3.00^\circ$ form. At the matched $3.77^\circ$ angle, the CDDA changes expected fuel by 0.1\% for the A319, $-0.5$\% for the B737-800, 7.7\% for the B767-400, and $-1.6$\% for the A340-300 relative to the CDA. Thus three of the four fully configured CDA designs match or beat the delayed-deceleration form at the same steep final where only the B767-400 retains a material architecture advantage. The steeper pairs also spend the stabilization budget. At $3.77^\circ$, the A319, B737-800, and A340-300 optima of both procedures sit at or near the 95\% floor, while both B767-400 optima remain essentially failure-free.

Fixed flap rules preserve the same-angle closeness. The architecture comparison does not depend on trigger optimization. With identical fixed trigger ladders at the same final angle and 5{,}000 ft capture altitude, the CDDA burns 1.6--8.4\% less than the CDA across the four airframes, both fixed rules, and all shared final angles. For the B737-800 at $3.77^\circ$, the minimum-speed rule burns 268.1 kg on the CDDA versus 275.5 kg on the CDA, and the midpoint rule burns 263.4 kg versus 278.5 kg. Every fixed-rule CDDA arm remains chance-feasible, although the A319 and A340-300 minimum-speed ladders at $3.77^\circ$ retain tail risk ($P = 0.955$ and $0.973$).

The category D maximum preserves the pattern. At $3.50^\circ$, the same-angle architecture gap is 0.3\% for the A319, 3.9\% for the B737-800, 1.1\% for the B767-400, and 2.5\% for the A340-300. Both $3.50^\circ$ architectures capture inside the standard glideslope service volume at 7.0--9.5 nautical mile, while the $3.00^\circ$ pairs capture beyond it at 11.0--12.0 nautical mile. Within the CDA alone, changing the final from $3.00^\circ$ to $3.50^\circ$ saves 11.9\% (A319), 17.5\% (B737-800), 12.1\% (B767-400), and 8.5\% (A340-300), which is most of the 12--22\% obtained by the full step to $3.77^\circ$. Publishing the $3.50^\circ$ Category D maximum therefore captures most of the angle benefit within current FAA design criteria for all four study airframes.

\subsection{The Headline Procedures and the Value of Trigger Optimization}
Table~\ref{tab:main} reports the two headline procedures, the conventional $3^\circ$ CDA and the $3.50^\circ$ CDDA, the steepest delayed-deceleration procedure publishable for all four airframes under current criteria, optimized under the identical formulation, wind model, and stabilization constraint, together with their two baselines. The two procedures differ in both the deceleration architecture and the final angle, so the savings here combine the two factors that Sec.~V.B separated, and they add the third: the value of optimizing the flap triggers within each procedure.

\begin{figure}[hbt!]
\centering
\begin{subfigure}{0.49\textwidth}
\includegraphics[width=\textwidth]{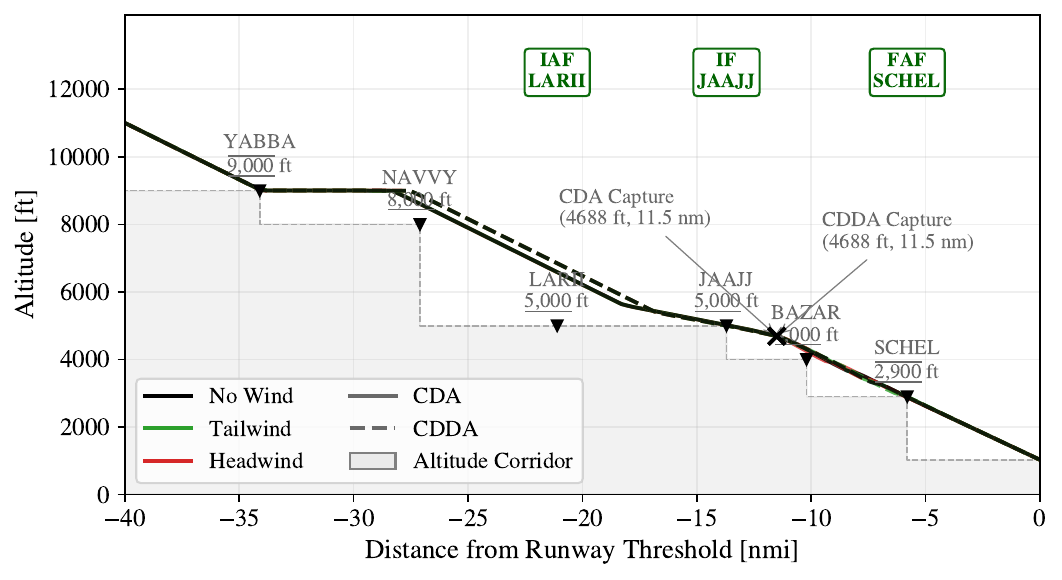}
\caption{B737-800 altitude.}
\end{subfigure}
\begin{subfigure}{0.49\textwidth}
\includegraphics[width=\textwidth]{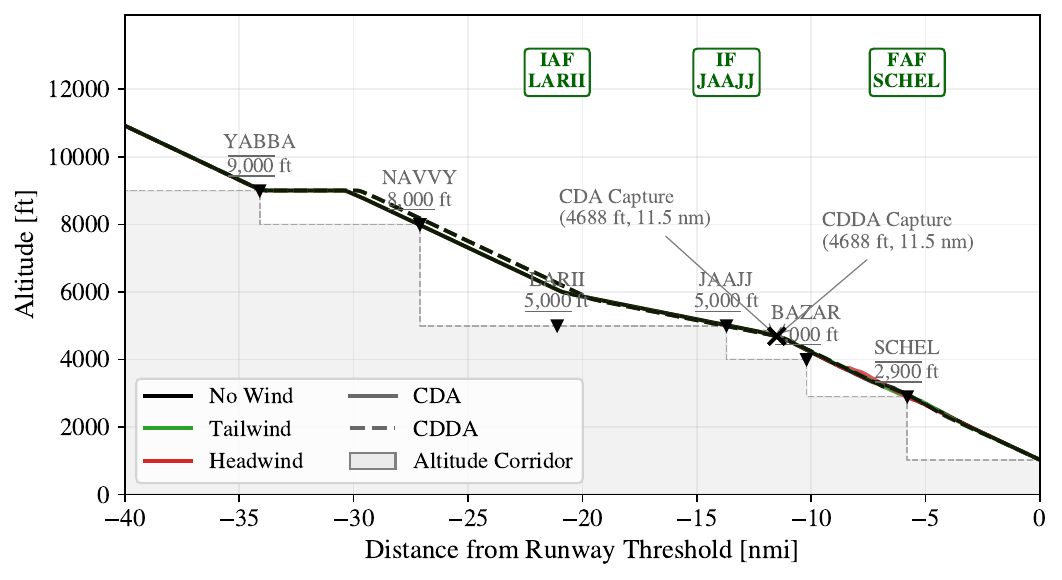}
\caption{A340-300 altitude.}
\end{subfigure}
\\
\begin{subfigure}{0.49\textwidth}
\includegraphics[width=\textwidth]{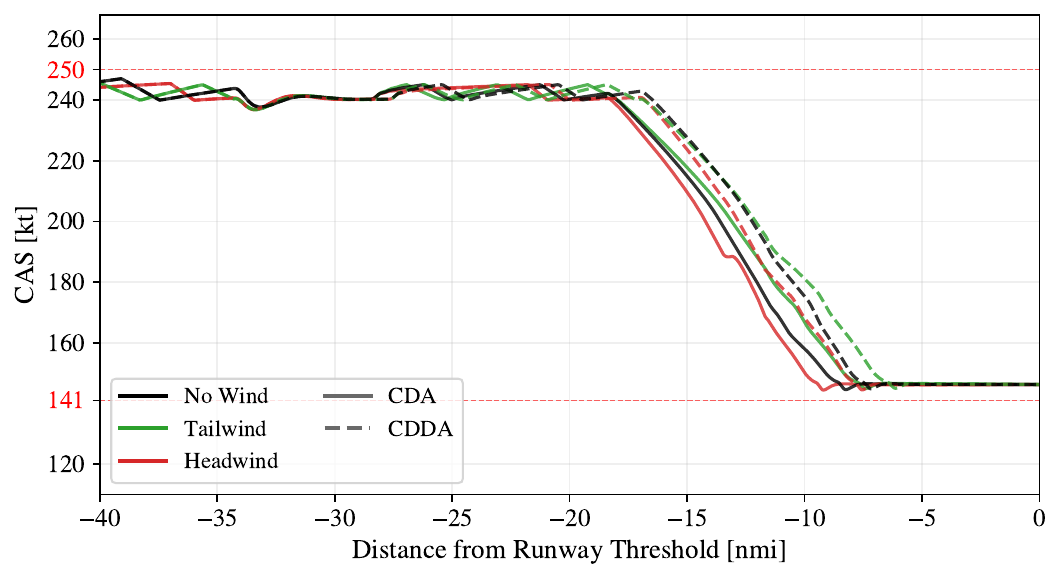}
\caption{B737-800 calibrated airspeed.}
\end{subfigure}
\begin{subfigure}{0.49\textwidth}
\includegraphics[width=\textwidth]{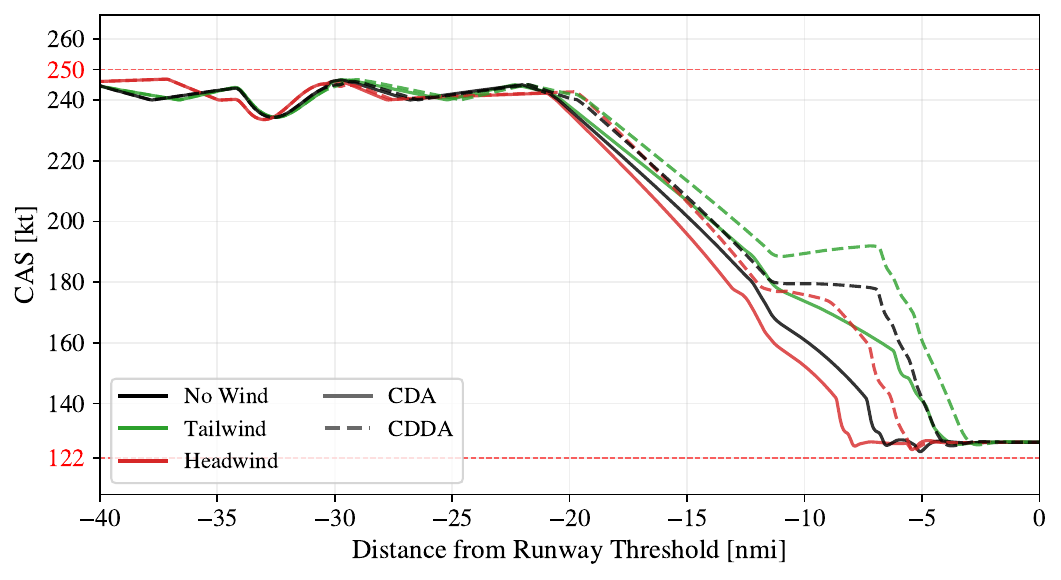}
\caption{A340-300 calibrated airspeed.}
\end{subfigure}
\caption{Optimized $3.00^\circ$ CDA versus optimized $3.00^\circ$ CDDA at no wind, tailwind, and headwind on altitude and calibrated airspeed.}
\label{fig:hist-alt}
\end{figure}

\begin{figure}[hbt!]
\centering
\begin{subfigure}{0.49\textwidth}
\includegraphics[width=\textwidth]{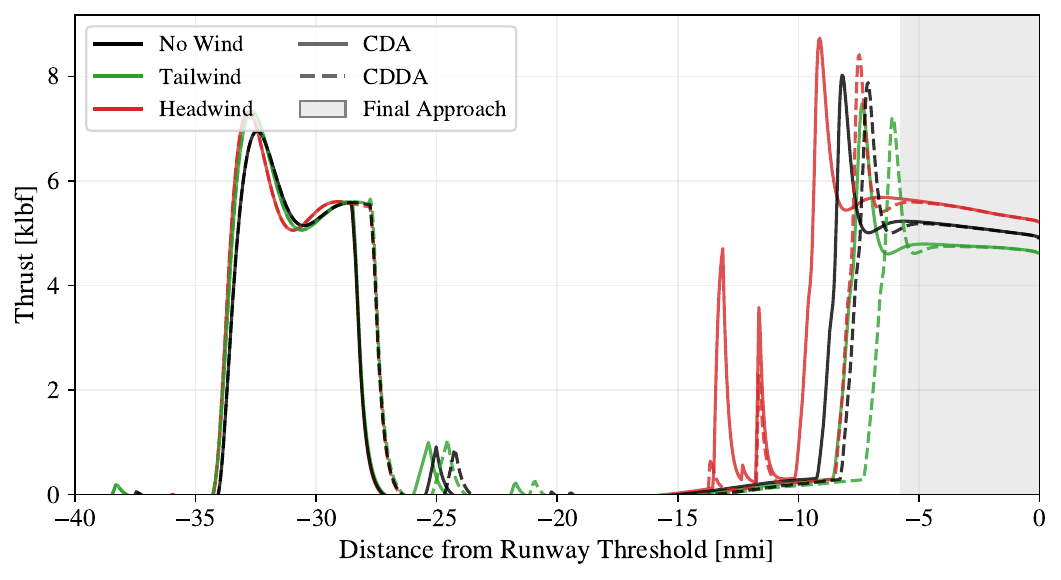}
\caption{B737-800 total thrust.}
\end{subfigure}
\begin{subfigure}{0.49\textwidth}
\includegraphics[width=\textwidth]{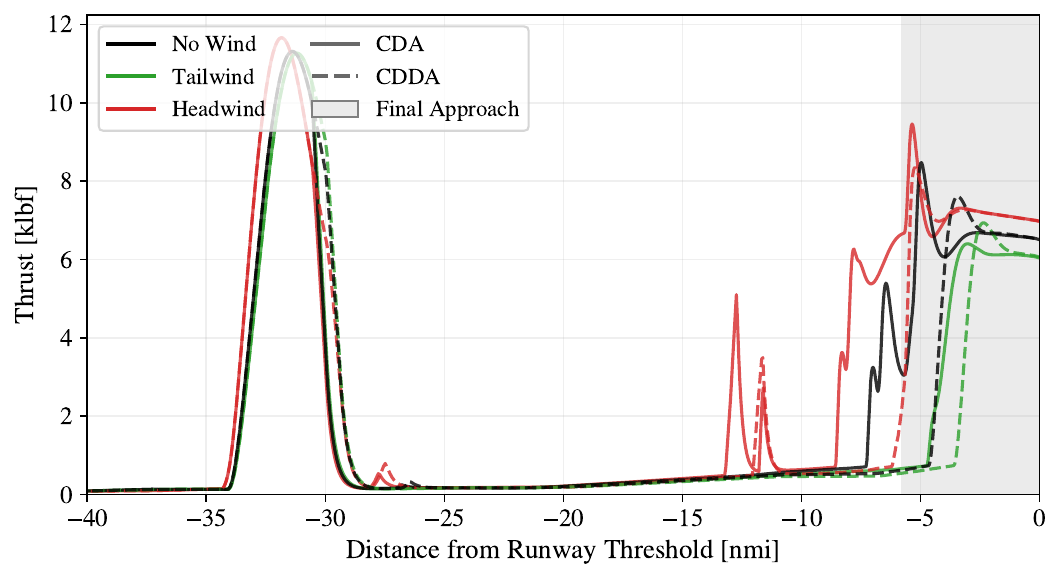}
\caption{A340-300 total thrust.}
\end{subfigure}
\\
\begin{subfigure}{0.49\textwidth}
\includegraphics[width=\textwidth]{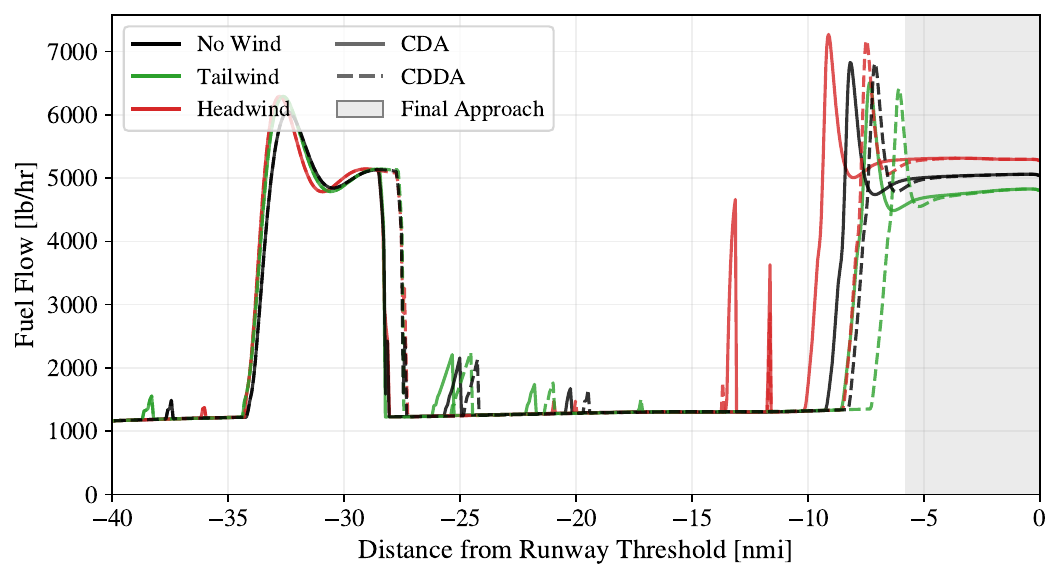}
\caption{B737-800 fuel flow.}
\end{subfigure}
\begin{subfigure}{0.49\textwidth}
\includegraphics[width=\textwidth]{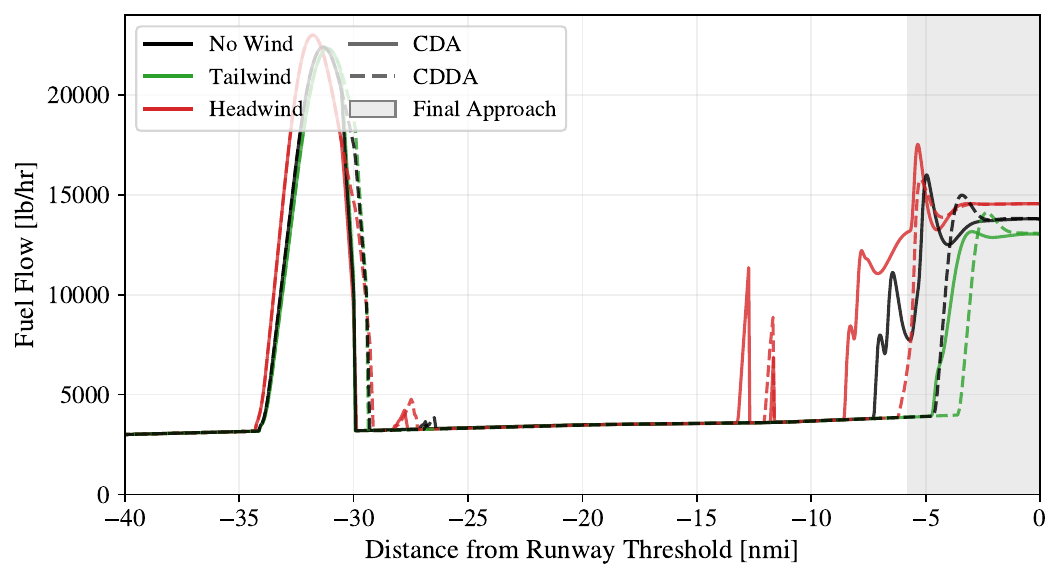}
\caption{A340-300 fuel flow.}
\end{subfigure}
\caption{Optimized $3.00^\circ$ CDA versus optimized $3.00^\circ$ CDDA at no wind, tailwind, and headwind on engine thrust and fuel-flow rate.}
\label{fig:hist-thrust}
\end{figure}

\begin{figure}[hbt!]
\centering
\begin{subfigure}{0.49\textwidth}
\includegraphics[width=\textwidth]{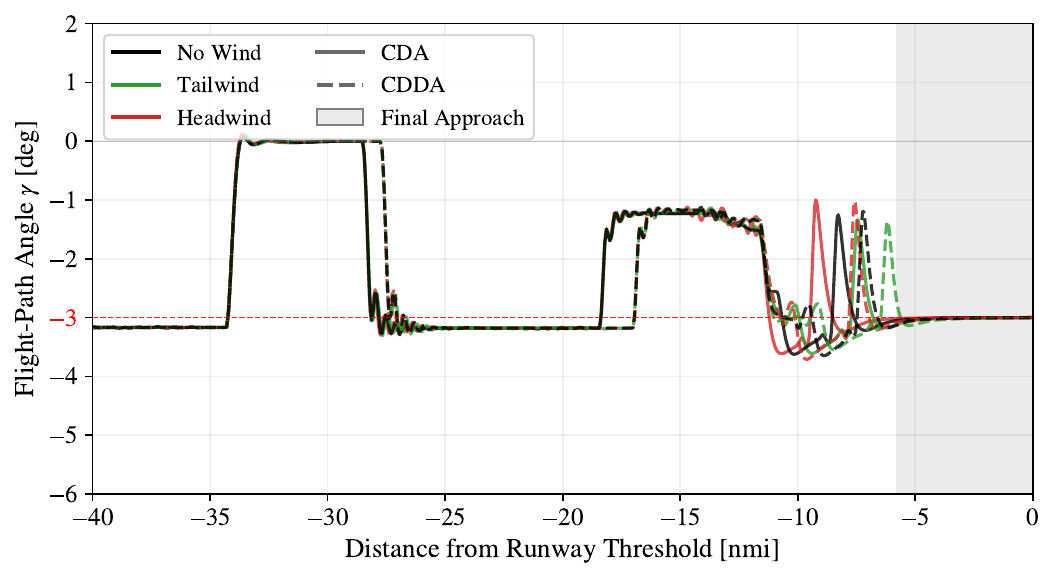}
\caption{B737-800 flight-path angle.}
\end{subfigure}
\begin{subfigure}{0.49\textwidth}
\includegraphics[width=\textwidth]{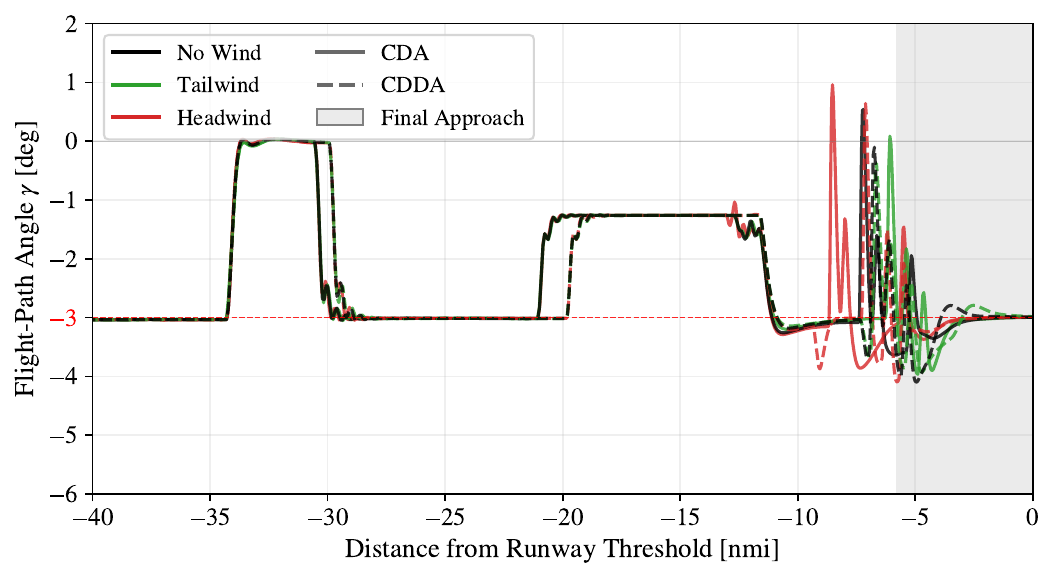}
\caption{A340-300 flight-path angle.}
\end{subfigure}
\\
\begin{subfigure}{0.49\textwidth}
\includegraphics[width=\textwidth]{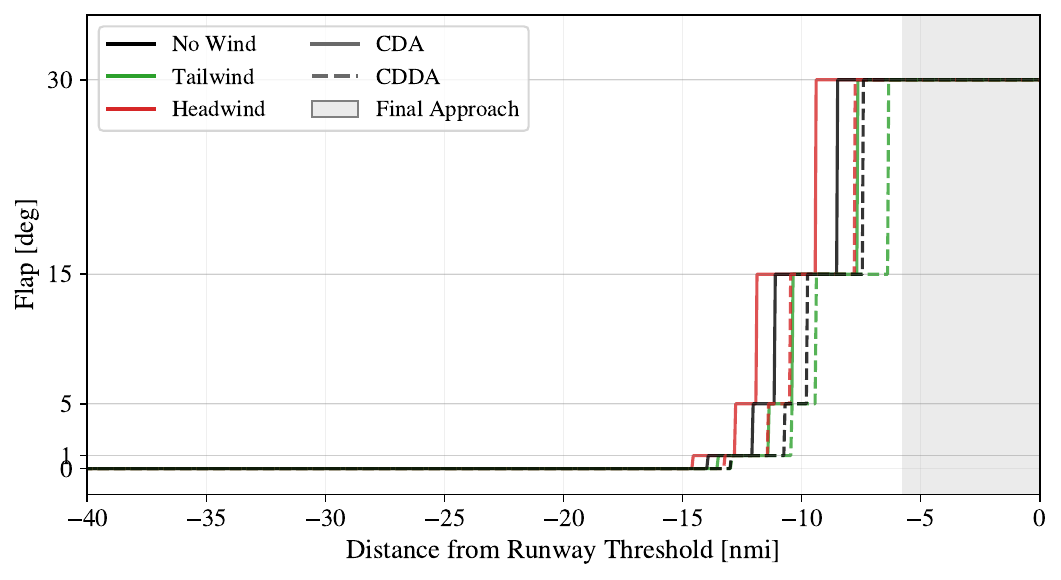}
\caption{B737-800 flap schedule.}
\end{subfigure}
\begin{subfigure}{0.49\textwidth}
\includegraphics[width=\textwidth]{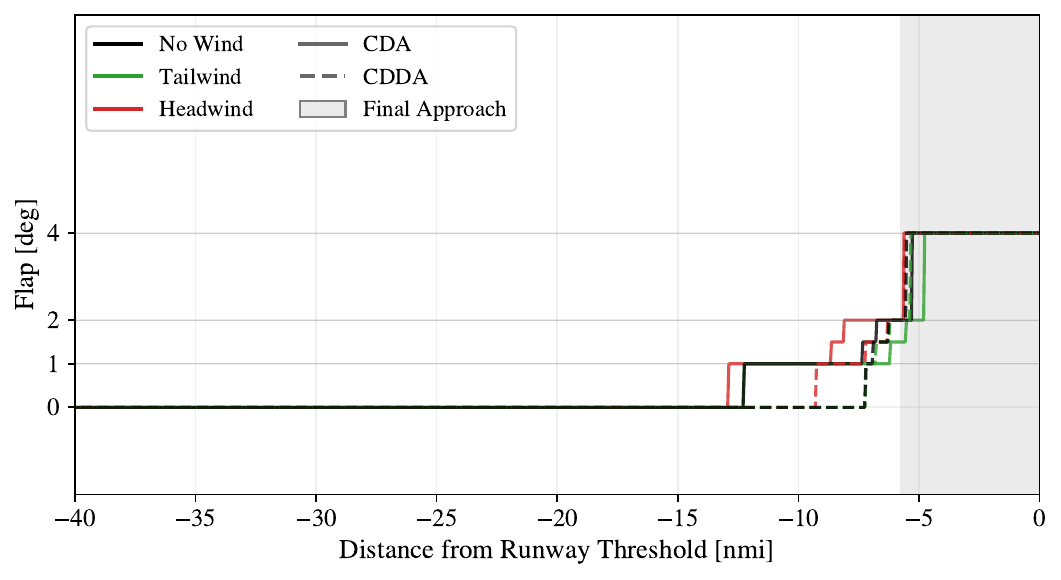}
\caption{A340-300 flap schedule.}
\end{subfigure}
\caption{Optimized $3.00^\circ$ CDA versus optimized $3.00^\circ$ CDDA at no wind, tailwind, and headwind on flight-path angle and flap schedule.}
\label{fig:hist-path}
\end{figure}

\begin{figure}[hbt!]
\centering
\begin{subfigure}{0.49\textwidth}
\includegraphics[width=\textwidth]{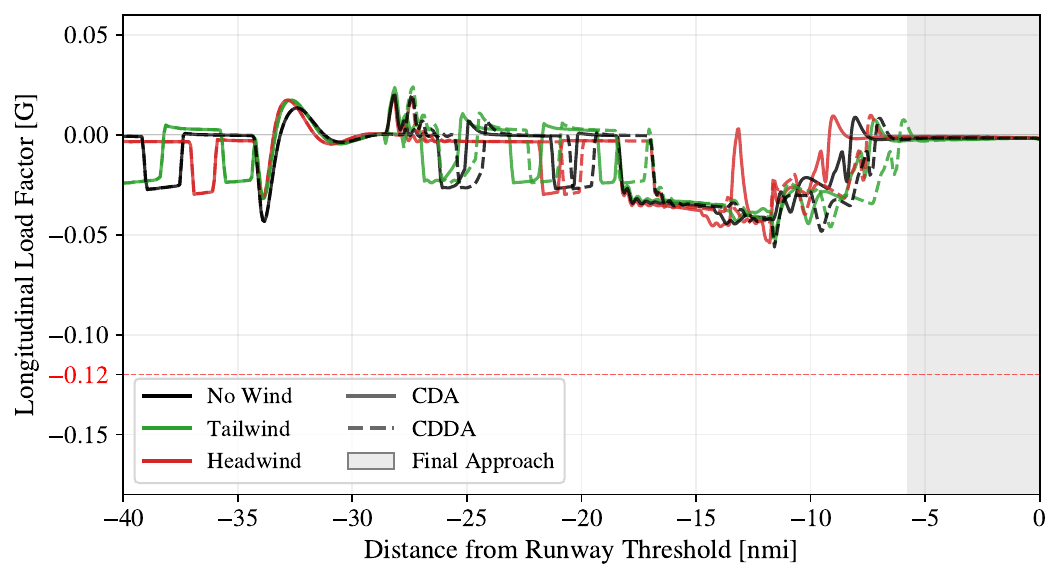}
\caption{B737-800 longitudinal load factor.}
\end{subfigure}
\begin{subfigure}{0.49\textwidth}
\includegraphics[width=\textwidth]{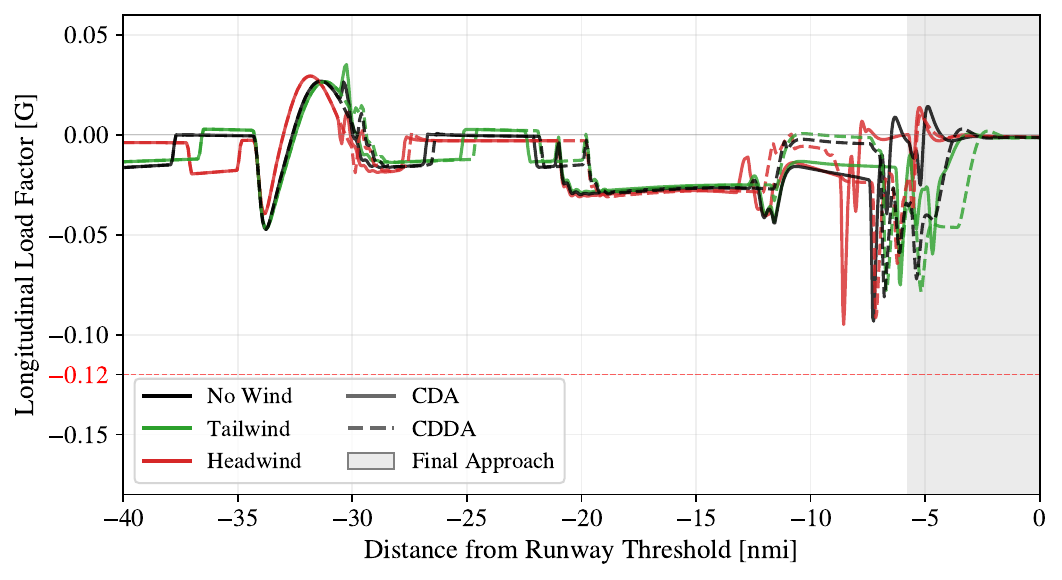}
\caption{A340-300 longitudinal load factor.}
\end{subfigure}
\\
\begin{subfigure}{0.49\textwidth}
\includegraphics[width=\textwidth]{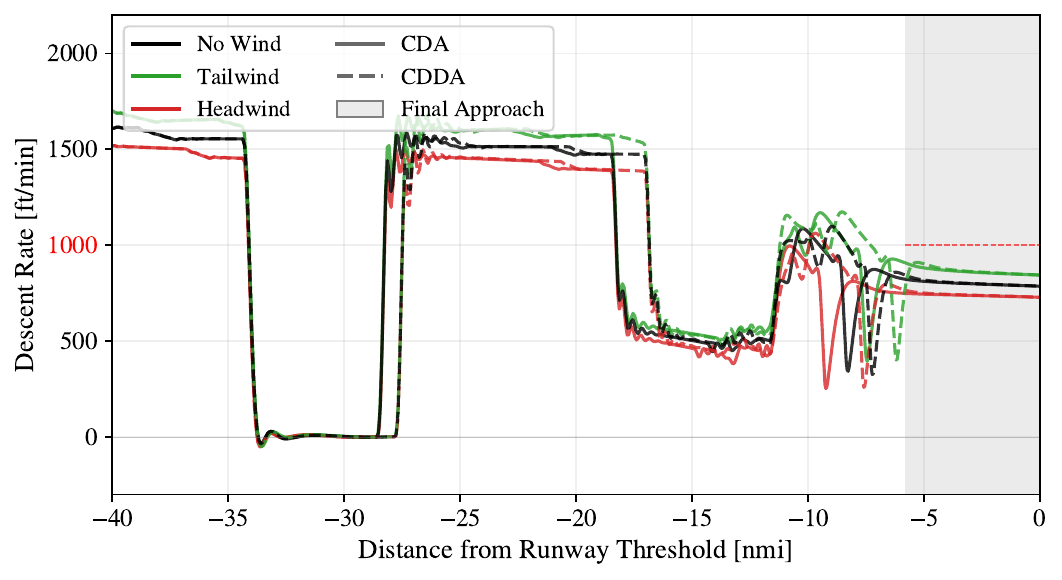}
\caption{B737-800 descent rate.}
\end{subfigure}
\begin{subfigure}{0.49\textwidth}
\includegraphics[width=\textwidth]{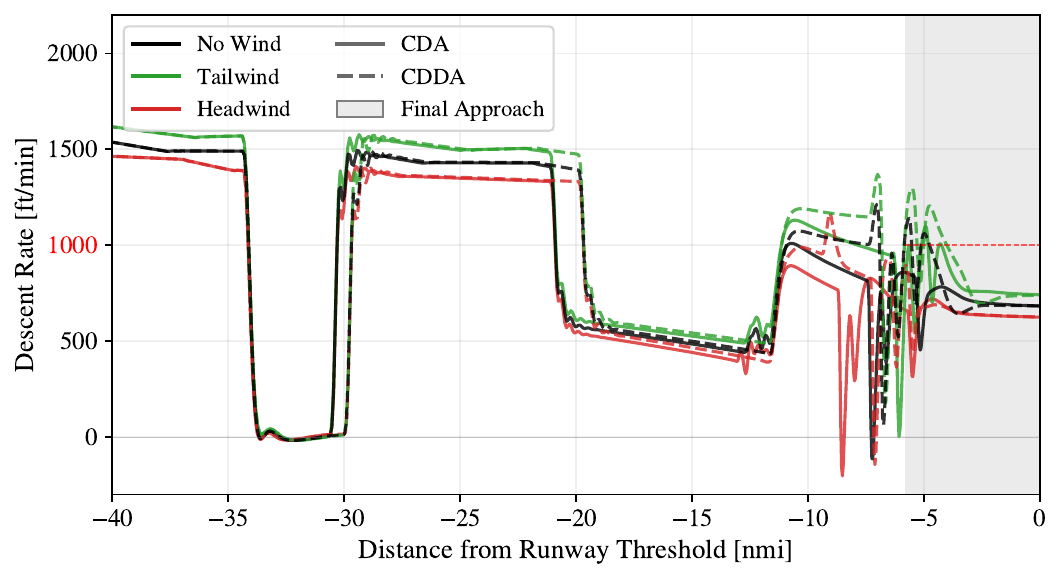}
\caption{A340-300 descent rate.}
\end{subfigure}
\caption{Optimized $3.00^\circ$ CDA versus optimized $3.00^\circ$ CDDA at no wind, tailwind, and headwind on longitudinal load factor and descent rate.}
\label{fig:hist-comfort}
\end{figure}

Figures~\ref{fig:hist-alt}--\ref{fig:hist-comfort} trace the optimized $3.00^\circ$ CDA and CDDA for the two detailed case-study airframes, the B737-800 and A340-300, at no wind and at $\pm 15$ kt wind. Every trace shown is stabilized; stronger tailwinds enter the accepted risk tail quantified in Sec.~V.D. In Figs.~\ref{fig:hist-thrust}--\ref{fig:hist-comfort}, the shaded band marks the final approach segment from the final approach fix to the runway threshold. The red dashed references mark the shared $3.00^\circ$ final in the flight-path-angle panels, the $-0.12$g deceleration limit of the stabilization constraint in the load-factor panels, and the 1{,}000 ft/min sink-rate element of the stabilized-approach criteria \cite{fsf2009stabilized} in the descent-rate panels. On the shared $3.00^\circ$ final, published floors force both architectures into the same capture geometry: both optima capture at 11.5 nautical mile, beyond the standard glideslope service volume, and their altitude histories are nearly indistinguishable. The speed and configuration schedules differ. The CDA begins decelerating upstream and crosses the capture point fully configured at approach speed. The CDDA holds the clean 240 kt descent speed several nautical mile deeper, shifts the shallow deceleration segments toward the runway, captures at landing-flap deployment speed, and completes configuration on final. Thrust and fuel flow remain at idle longer on the CDDA, producing the same-angle saving in Table~\ref{tab:ablation}. At no wind, that saving is 7.8 kg for the B737-800 and 62.3 kg for the A340-300. The longitudinal load factor stays within the $-0.12$g stabilization limit for both architectures under all three winds. Because the two procedures share glideslope and approach speed, their final-descent sink rates are nearly identical and remain below 1{,}000 ft/min; the steeper-final sink-rate cost is quantified in finding (e).

\begin{table}[hbtp!]
\caption{\label{tab:main} Optimized designs and baselines under the common stochastic formulation. The conventional $3.00^\circ$ CDA and the $3.50^\circ$ CDDA, both publishable for all four airframes under current FAA glidepath criteria. Baselines are flown at the corridor's current 5{,}000 ft capture altitude (12.48 nautical mile on the $3.00^\circ$ final, 10.69 on the $3.50^\circ$). The edge-of-criteria $3.77^\circ$ optima appear in Table~\ref{tab:ablation}.}
\centering
\scriptsize
\setlength{\tabcolsep}{2.5pt}
\begin{tabularx}{\textwidth}{llc>{\raggedright\arraybackslash}Xcc c>{\raggedright\arraybackslash}Xcc}
\hline
 &  & \multicolumn{4}{c}{CDDA, $3.50^\circ$ final} & \multicolumn{4}{c}{CDA, $3.00^\circ$ final} \\
\cline{3-6}\cline{7-10}
Aircraft & Rule & $d_{\mathrm{cap}}$ & Trigger speeds, kt & $E[f]$ & $P_s$ & $d_{\mathrm{cap}}$ & Trigger speeds, kt & $E[f]$ & $P_s$ \\
\hline
A319 & Optimized & 9.5 & 170, 148, 148, 142, 125 & 169.7 & 0.955 & 11.5 & 190, 131, 131, 128, 108 & 193.3 & 0.985 \\
 & Minimum-speed & 10.69 & 200, 159, 141, 141, 141 & 182.2 & 0.993 & 12.48 & 200, 159, 141, 141, 141 & 219.7 & 1.000 \\
 & Midpoint & 10.69 & 190, 166, 166, 157, 142 & 189.1 & 0.996 & 12.48 & 190, 166, 166, 157, 142 & 230.8 & 1.000 \\
\hline
B737-800 & Optimized & 7.0 & 220, 220, 150, 150, 150 & 233.0 & 0.980 & 11.5 & 190, 180, 150, 150, 149 & 293.8 & 1.000 \\
 & Minimum-speed & 10.69 & 220, 200, 190, 160, 160 & 280.0 & 1.000 & 12.48 & 220, 200, 190, 160, 160 & 318.1 & 1.000 \\
 & Midpoint & 10.69 & 210, 202, 172, 172, 167 & 274.9 & 1.000 & 12.48 & 210, 202, 172, 172, 167 & 321.0 & 1.000 \\
\hline
B767-400 & Optimized & 9.0 & 217, 190, 177, 163, 147 & 482.5 & 1.000 & 11.0 & 217, 217, 217, 208, 147 & 555.1 & 1.000 \\
 & Minimum-speed & 10.69 & 233, 213, 193, 173, 173 & 533.2 & 1.000 & 12.48 & 233, 213, 193, 173, 173 & 605.6 & 1.000 \\
 & Midpoint & 10.69 & 232, 212, 192, 189, 165 & 532.1 & 1.000 & 12.48 & 232, 212, 192, 189, 165 & 607.1 & 1.000 \\
\hline
A340-300 & Optimized & 9.0 & 167, 152, 143, 143, 122 & 669.0 & 0.973 & 11.5 & 179, 131, 126, 126, 126 & 750.4 & 1.000 \\
 & Minimum-speed & 10.69 & 185, 153, 141, 136, 136 & 705.7 & 1.000 & 12.48 & 185, 153, 141, 136, 136 & 826.5 & 1.000 \\
 & Midpoint & 10.69 & 192, 173, 161, 161, 151 & 759.7 & 1.000 & 12.48 & 192, 173, 161, 161, 151 & 896.2 & 1.000 \\
\hline
\end{tabularx}
\end{table}

From Table~\ref{tab:main}, we draw five major findings and list them here.

\emph{(a) The compliant steep-final package saves 11--21\%, mostly through glideslope angle.} Comparing the two headline procedures, the $3.50^\circ$ CDDA burns 12.2\% less expected fuel than the conventional $3^\circ$ CDA for the A319 (169.7 vs 193.3 kg), 20.7\% less for the B737-800 (233.0 vs 293.8 kg), 13.1\% less for the B767-400 (482.5 vs 555.1 kg), and 10.8\% less for the A340-300 (669.0 vs 750.4 kg). Relative to the $3^\circ$ CDA flown with the minimum-speed rule, the optimized CDDA saves 22.8\% (A319), 26.8\% (B737-800), 20.3\% (B767-400), and 19.1\% (A340-300). The same-angle factorial of Sec.~V.B attributes most of this saving to final angle where an optimized CDA on the same $3.50^\circ$ final reduces the architecture gap to 0.3--3.9\%.

\emph{(b) Optimized triggers beat both fixed rules and spend the risk budget.} Against the minimum-speed rule, optimized CDDA triggers save 6.9\% (A319), 16.8\% (B737-800), 9.5\% (B767-400), and 5.2\% (A340-300). Against the midpoint rule, they save 10.3\%, 15.2\%, 9.3\%, and 11.9\%. The fixed rules remain chance-feasible on every airframe ($P = 0.993$--$1.000$), but burn more fuel. The optima use the available stabilization margin. The A319 design sits at $P = 0.955$, the A340-300 and B737-800 at 0.973 and 0.980, and the B767-400 at 1.000. For the A319, tailwinds beyond roughly $+17$ kt, with about 4.5\% probability, lead to a go-around. Section~V.D quantifies this trade.

\emph{(c) Optimized trigger ladders are structured rather than uniform.} The B737-800 optimum extends the first two detents early (220 kt), adding partial-flap drag when the late deceleration needs it, and holds the deep detents late (150 kt). The B767-400 grades its ladder from 217 kt down to a late 147 kt landing flap. The Airbus optima set every trigger low, 170--125 kt for the A319 and 167--122 kt for the A340-300, using cleanliness and inertia for deceleration. These patterns are consistent with the flight-data correlation between flaps-extended time and fuel \cite{dumont2012fuel}.

\emph{(d) The steeper final brings capture inside the service volume.} Every optimized $3.50^\circ$ CDDA captures the glideslope at 7.0--9.5 nautical mile, inside the 10 nautical mile standard glideslope service volume \cite{faa_aim}. The $3^\circ$ CDA cannot. The published 5{,}000 ft altitude floors force capture to 11.0--11.5 nautical mile, outside the standard volume. The same repair appears for the CDA when it is optimized on the $3.50^\circ$ final, where it captures at 7.0--9.0 nautical mile. The geometry improvement is therefore an angle effect, not a delayed-deceleration effect.

\emph{(e) Sink rate is the operational cost of the steeper final.} Each optimized arm was audited on the 51-node verification grid against the 1{,}000 ft/min sink-rate element of Sec.~IV.C. On the $3.00^\circ$ final, the element is met almost everywhere: zero-wind gate-crossing sink rates are 660--830 ft/min, and only the A319 designs approach the limit in the strongest tailwinds. At $3.50^\circ$, calm-air sink rates remain compliant for every airframe, 772--969 ft/min at the gate, with exceedances confined to tailwinds. At $3.77^\circ$, the element binds for faster aircraft, where the B767-400 requires 1{,}035--1{,}045 ft/min in calm air and exceeds 1{,}000 ft/min with probability about 0.8, and the B737-800 exceeds the limit from a $+3$ kt tailwind onward. The level-deceleration DDA follows the same pattern, confirming that sink rate follows from angle and approach speed rather than architecture. Under industry criteria, these approaches are not prohibited, but exceedances require the special briefing for approaches demanding more than 1{,}000 ft/min \cite{fsf2009stabilized}. The $3.50^\circ$ headline procedure therefore retains most of the angle benefit while limiting special-briefing exposure; the $3.77^\circ$ cases bound the additional benefit available under more automated future operations.

\subsection{Robustness Sensitivity: The Role of the DDA's Level Deceleration Segment}
The robustness comparison uses the $3.77^\circ$ final, where delayed deceleration is most stressed and the level segment has the clearest role. The optimized $3.77^\circ$ CDDA designs for the A319, B737-800, and A340-300 have stabilization probabilities of 0.955, nearly exhausting the 5\% risk budget. The unstabilized cases occur in the tailwind band beyond roughly $+17$ kt, about 4.5\% of the truncated-normal wind mass. There the aircraft crosses the 1{,}000 ft gate faster than $V_{\mathrm{REF}}+15$ kt and goes around, as permitted by the chance constraint. In the strongest tailwinds, the failure can become a deceleration deadlock where idle thrust on the steep final cannot slow the clean airframe below the next flap placard limit, so configuration waits until deployment becomes legal.

To quantify what that risk budget is worth, every arm was re-optimized with $\varepsilon = 0$ and the design must stabilize at every one of the 51 fine-grid wind nodes. Table~\ref{tab:robust} reports the resulting minima for all four airframes, alongside the level-deceleration DDA optimized under the identical formulation.

\begin{table}[hbt!]
\caption{\label{tab:robust} Robustness sensitivity at the $3.77^\circ$ final ($3.00^\circ$ for the CDA rows). Expected fuel (kg) of each optimized arm at the 5\% chance budget and under a zero-failure requirement over all 51 wind nodes. Bold marks the preferred delayed-deceleration variant at each risk level.}
\centering
\begin{tabular}{llcc}
\hline
Aircraft & Procedure & $\varepsilon = 5\%$ & $\varepsilon = 0$ (zero failures) \\
\hline
A319 & CDA & 193.3 & 197.3 \\
 & DDA & 167.4 & \textbf{171.0} \\
 & CDDA & \textbf{166.1} & Infeasible (best $P = 0.980$) \\
\hline
B737-800 & CDA & 293.8 & 293.8 \\
 & DDA & 230.1 & \textbf{235.1} \\
 & CDDA & \textbf{229.1} & 243.1 \\
\hline
B767-400 & CDA & 555.1 & 555.1 \\
 & DDA & 499.2 & 499.2 \\
 & CDDA & \textbf{438.6} & \textbf{445.4} \\
\hline
A340-300 & CDA & 750.4 & 750.4 \\
 & DDA & 681.4 & 700.1 \\
 & CDDA & \textbf{669.4} & \textbf{691.8} \\
\hline
\end{tabular}
\end{table}

\begin{figure}[hbt!]
\centering
\begin{subfigure}{0.49\textwidth}
\includegraphics[width=\textwidth]{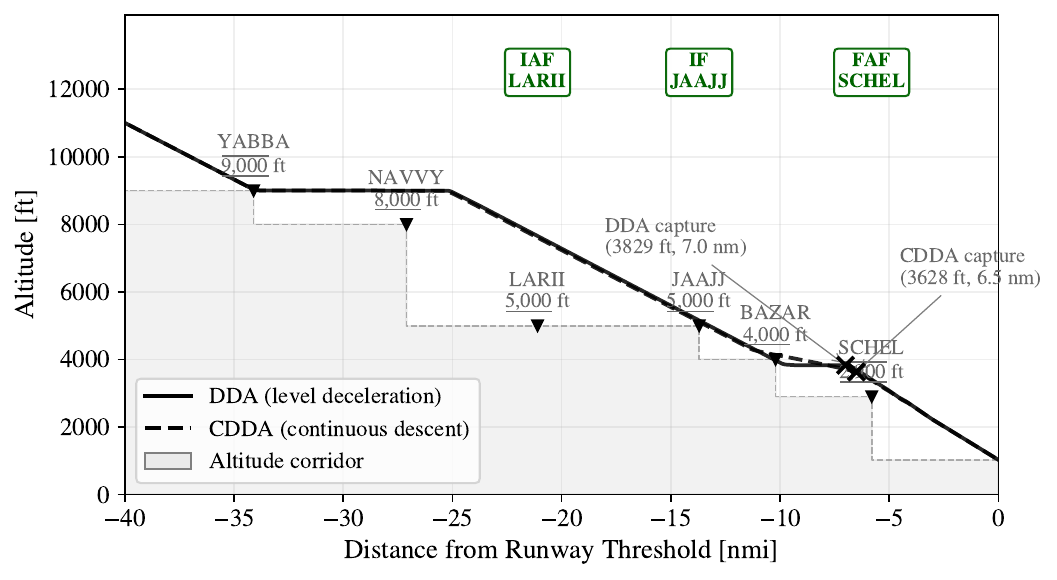}
\caption{B737-800 descent path.}
\end{subfigure}
\begin{subfigure}{0.49\textwidth}
\includegraphics[width=\textwidth]{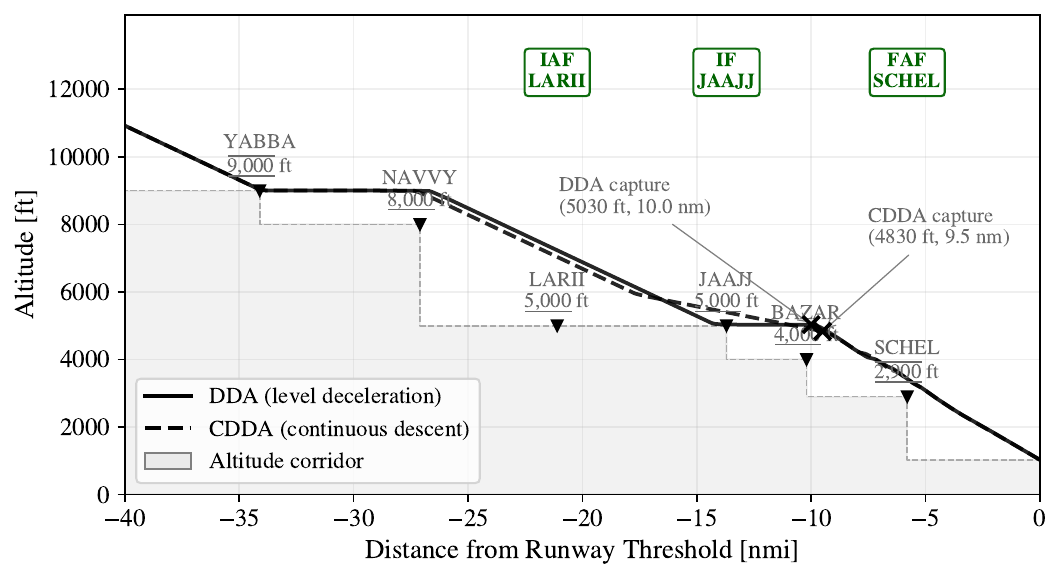}
\caption{A340-300 descent path.}
\end{subfigure}
\caption{The two delayed-deceleration variants at their chance-constrained optima. The DDA decelerates on a level segment before capture, and the CDDA descends through to a lower and closer capture.}
\label{fig:ddavar}
\end{figure}

Table~\ref{tab:robust} supports three conclusions. \emph{The level deceleration segment is the DDA's tailwind-robustness buffer, and its value depends on airframe and risk tolerance.} At the 5\% budget, removing the level segment saves 0.4\% expected fuel for the B737-800, 0.8\% for the A319, 1.8\% for the A340-300, and 12.1\% for the B767-400. Both variants sit near the chance boundary for the three lighter airframes ($P = 0.955$--0.965), so the robustness value of the level segment becomes most visible when the risk budget is removed. Figure~\ref{fig:ddavar} shows the mechanism. Descending deceleration crosses the published floors higher and captures lower and closer, but removes the level segment on which a tailwind-fast aircraft can finish decelerating.

\emph{Demanding zero failures changes the preferred delayed-deceleration variant by weight class.} The light, aerodynamically clean A319 has no CDDA design that stabilizes at every wind node; the best fine-verified design achieves $P = 0.980$. Its robust delayed-deceleration procedure is therefore the level-deceleration DDA at 171.0 kg. The B737-800 also reverts to the level segment, with 235.1 kg for the robust DDA versus 243.1 kg for the robust CDDA. The two wide-bodies keep the continuous-descent form, 445.4 kg versus 499.2 kg for the B767-400 and 691.8 kg versus 700.1 kg for the A340-300. The lighter and cleaner the airframe, the more it benefits from the level segment under a zero-failure requirement.

\emph{The attribution is insensitive to the risk budget.} Even under the zero-failure requirement, the best delayed-deceleration variant beats the optimized $3^\circ$ CDA by 13.3\% (A319), 20.0\% (B737-800), 19.8\% (B767-400), and 7.8\% (A340-300). A $3.77^\circ$ CDA re-optimized under the same zero-failure requirement closes the same-angle gap to $+1.1$\%, $-0.3$\%, $+6.2$\%, and $-0.1$\%. The risk budget changes which delayed-deceleration variant is preferred and slightly changes fuel, but the main attribution to glideslope angle remains.

\subsection{Limitations}
The stochastic model reduces wind to one along-corridor component with a fixed shear profile. It does not represent crosswinds, gust-scale turbulence, wind forecast error relative to a nonzero forecast, or day-to-day variation of the shear exponent. Landing mass is fixed per airframe, pilot response time is nominal, and the corridor is a single straight-in geometry. The steep final also carries operational restrictions outside the fuel comparison. A glidepath steeper than about $3.25^\circ$ lies outside current autoland certifications, so the $3.50^\circ$ and $3.77^\circ$ finals are manually flown category I operations, and the flight demonstration used a test glideslope at the FAA William J. Hughes Technical Center rather than a published procedure \cite{thomas2021modeling}. Establishing such a procedure for fuel rather than obstacle-clearance reasons would require Flight Standards approval \cite{faa_terps}. The same formulation can incorporate additional uncertainty dimensions where mass and pilot delay are the most direct next additions from prior RNAV separation analysis \cite{ren2007modeling, kendall2020stochastic}.

\section{Conclusion}
This paper formulated approach-procedure design as a chance-constrained stochastic optimization. The design variables are glideslope-capture distance and flap-deployment trigger speeds. The objective is expected fuel over a wind distribution, and the constraint is a 95\% stabilized-approach probability. A fast-time aircraft-and-FMS simulation evaluates fuel and stabilization for each candidate. Applying the same formulation to the CDA and the CDDA across two narrow-body and two wide-body airframes, and re-optimizing both at $3.00^\circ$, $3.50^\circ$, and $3.77^\circ$, separates deceleration architecture from glideslope angle. The main result is straightforward. Glideslope angle and optimized flap schedule drive the fuel saving, while deceleration architecture is the weakest of the three factors. The results support five findings,
\begin{itemize}
\item \emph{Deceleration Architecture.} At matched glideslope the two architectures deliver similar expected fuel. Delayed deceleration accounts for 2.6--7.7\% at $3.00^\circ$, 0.3--3.9\% at $3.50^\circ$, and $-2$ to $+8$\% at $3.77^\circ$, material only for the B767-400.
\item \emph{Glide Path Angle.} The steeper final moves glideslope capture inside the standard service volume that $3^\circ$ procedures violate on this corridor. At the $3.50^\circ$ Category D maximum, the CDA captures at 7.0--9.0 nautical mile and burns 8--18\% less than its $3^\circ$ form, recovering most of the 12--22\% obtained at $3.77^\circ$ while remaining within current design criteria for all study airframes.
\item \emph{Flap Schedule.} Within every architecture--angle pair, optimized trigger speeds beat both fixed flap rules by 2--17\%. The optimal ladders are structured. Partial flaps deploy where their drag aids deceleration, and deep flaps and gear are delayed as much as stabilization allows.
\item \emph{Risk Budget.} The stochastic constraint affects the optimum. Fuel-optimal CDDA designs use most of the risk budget, with stabilization probabilities of 0.955--0.980 for three of the four airframes at $3.50^\circ$ and at the 0.955 boundary at $3.77^\circ$.
\item \emph{Airframe and Risk Budget.} The level deceleration segment is a tailwind-robustness buffer whose value depends on aircraft and risk tolerance. Under a zero-failure requirement, the A319 has no feasible level-segment-free CDDA, the B737-800 prefers the level-deceleration DDA, and both wide-bodies prefer the CDDA.
\end{itemize}

Three results extend beyond this corridor, (i) the CDDA architecture, which combines continuous descent with delayed deceleration; (ii) the stochastic optimization algorithm for flap-deployment speeds under wind uncertainty; and (iii) the attribution result that glideslope angle and flap schedule dominate the choice between early and delayed deceleration. The future works include additional uncertainty dimensions, multiple corridor geometries, and joint fuel-noise objectives within the same quadrature framework. Operational extensions include wind-conditional procedure variants selected through the automatic terminal information service and onboard flap triggers re-optimized against an uplinked descent forecast.

\section*{Acknowledgments}
This work was supported by the National Aeronautics and Space Administration (NASA) University Leadership Initiative (ULI) program under project ``Autonomous Aerial Cargo Operations at Scale,'' via grant No.~80NSSC21M071 to the University of Texas at Austin. Any opinions, findings, conclusions, or recommendations expressed in this material are those of the authors and do not necessarily reflect the
views of the project sponsor. The authors also thank Dr. Jiacheng Xie of the Georgia Institute of Technology, an instrument-rated pilot, and Jim Allerdice, the chief designer of the Area Navigation (RNAV) infrastructure at A80 TRACON, for helpful discussions regarding charted procedures at KATL.

\bibliography{ref}

\end{document}